\documentclass[aps,pra,amsmath,amssymb,floatfix,twocolumn,amsmath,superscriptaddress,twocolumn,nofootinbib,tighten,letterpaper]{revtex4-2}
\usepackage{multirow}
\usepackage{bbold}

\usepackage{color}
\usepackage{mathrsfs}
\usepackage{hyperref}
\usepackage[normalem]{ulem}
\usepackage{bm}

\usepackage[caption=false]{subfig} 

\usepackage{stackrel}

\usepackage{amsfonts, relsize, color}
\usepackage{graphics}
\usepackage{graphicx}

\usepackage[table]{xcolor}

\usepackage{hyperref}
\usepackage{color}
\usepackage{comment}

\definecolor{RowColor}{rgb}{0.92,0.92,1.0}

\renewcommand\vec[1]{\ensuremath\boldsymbol{#1}} 

\begin{document}
\title{Nodal pair density waves from a quarter-metal in crystalline graphene multilayers}

\author{Sk Asrap Murshed}
\affiliation{Department of Physics, Lehigh University, Bethlehem, Pennsylvania, 18015, USA}

\author{Bitan Roy}~\thanks{Contact author: bitan.roy@lehigh.edu}
\affiliation{Department of Physics, Lehigh University, Bethlehem, Pennsylvania, 18015, USA}

\date{\today}
\begin{abstract}
Crystalline graphene heterostructures, namely, Bernal bilayer graphene (BBLG) and rhombohedral trilayer graphene (RTLG), for example, subject to perpendicular electric displacement fields, display a rich confluence of competing orders, resulting in a valley-degenerate, spin-polarized half-metal at moderate doping, and a spin- and valley-polarized (non-degenerate) quarter-metal at lower doping. Here we show that such a quarter-metal can be susceptible toward the nucleation of a unique spin- and valley-polarized superconducting ground state, accommodating \emph{odd-parity} (dominantly $p$ wave in BBLG and $f$ wave in RTLG) inter-layer Cooper pairs that break the translational symmetry, giving rise to a Kekul\'e (in BBLG) or columnar (in RTLG) pair density wave. Due to the trigonal warping in the normal state, the superconducting ground state produces three-fold rotationally symmetric isolated Fermi rings of normal fermions, which can manifest via linear in temperature scaling of the specific heat. We present scaling of the zero-temperature pairing amplitude and the transition temperature of such pair density wave in the presence of trigonally warped disconnected, annular, and simply connected Fermi rings in the normal state, subject to an effective attractive interaction within a mean-field approximation.          
\end{abstract}

\maketitle

\section{Introduction}

At the dawn of the 20th century, a series of pioneering experiments led by Onnes unearthed a novel quantum phase of matter in ordinary metals that shows zero electrical resistance below a few kelvins: superconductors. Almost half a century later, a successful microscopic theory of superconductivity by Bardeen, Cooper, and Schrieffer, spurred a new wave of scientific investigation in this direction~\cite{schieffer:book, tinkham:book, BCS:original}. Since then theoretical proposals and (possible) material realizations of high-$T_c$, nodal, time-reversal symmetry-breaking, mixed-parity, and topological superconductors have enriched this landscape~\cite{uedasigrist:review, hightc:review, vovovik:book, tsc:review}. From this jammed crowd, one member takes the center stage, Fulde-Ferrell-Larkin-Ovchinnikov (FFLO) superconductors, as the constituting Cooper pairs break the translational symmetry therein~\cite{fulde-ferrell, larkin-ovchinikov}. When the periodic modulation of the Cooper amplitude becomes commensurate with the underlying enlarged lattice periodicity, a special class of FFLO pairing emerges: pair density wave (PDW)~\cite{roy-herbut:PDW, cuprate:PDW, TMDC:PDW, raghu:PDW, hongyao:PDW, pnictide:PDW, kagome:PDW, otherPDW:1, otherPDW:2, otherPDW:3, otherPDW:4}.

\begin{figure}[b!]
\centering
\includegraphics[width=0.95\linewidth]{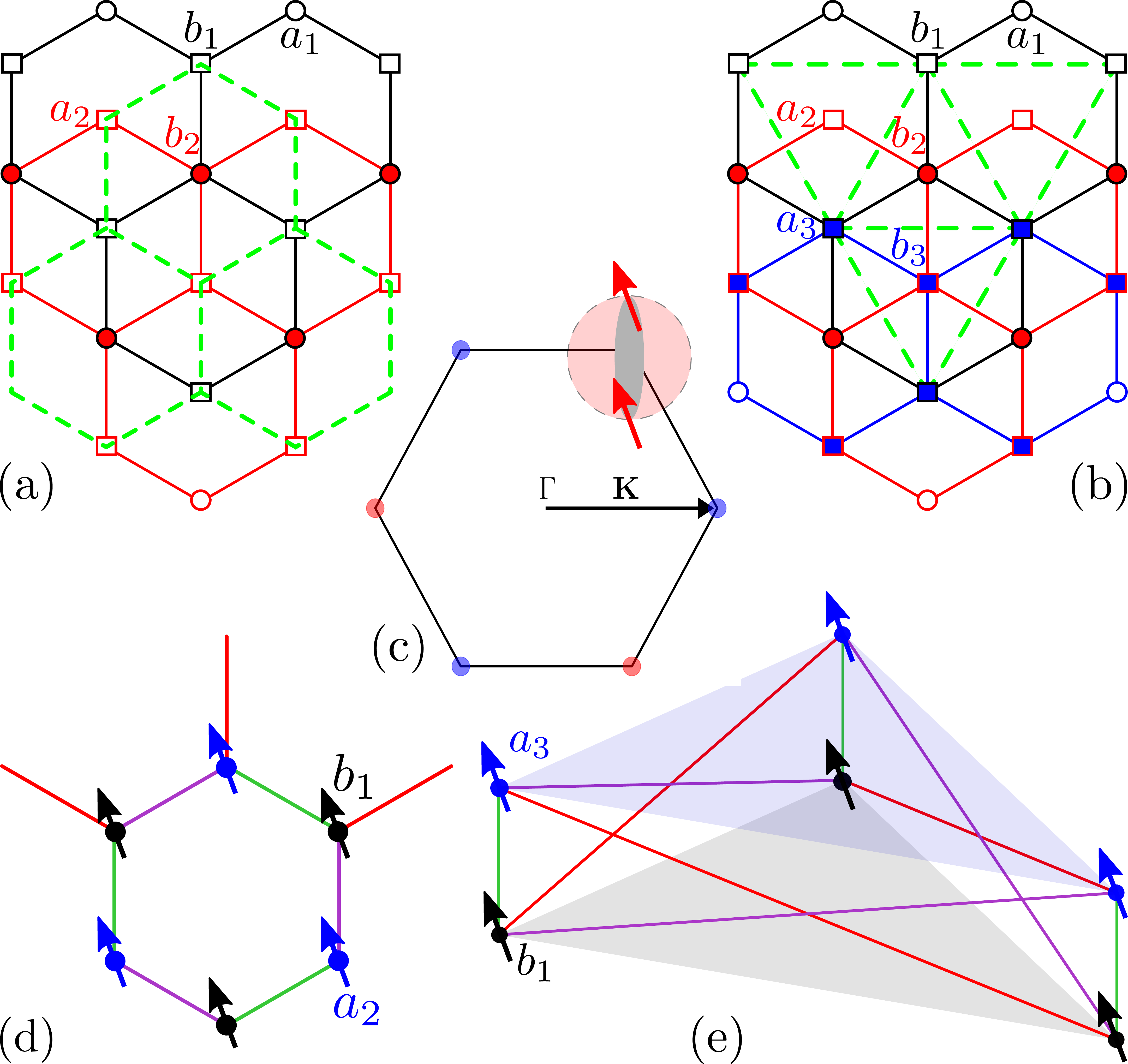}
\caption{Top view of (a) BBLG and (b) RTLG. Split-off bands at high energies reside predominantly on the overlapping sites, namely, $a_1$, $b_2$ in BBLG and $a_1$, $b_2$ and $a_2$, $b_3$ in RTLG. Integration over the high-energy bands yields an effective low-energy two-band model on two sets of sites, which form an effective honeycomb (prismlike) lattice in BBLG (RTLG), highlighted in green. (c) PDW Cooper pair (gray) around a Fermi pocket (pink) near one particular valley in QM. (d) Real space structure of PDW Cooper pairs on these two lattices are shown in (d) and (e), respectively, which along the green line is $\Delta \cos \alpha$, and along the red and blue lines is $\Delta \cos ( \alpha \pm 2\pi/3)$, respectively. Here $\Delta$ ($\alpha$) is the overall amplitude (internal angle) of the pairing.
}\label{fig:lattice}
\end{figure}

Nodal superconductors are fascinating as they harbor gapless neutral Bogoliubov-de Gennes (BdG) quasiparticles, typically around point or line nodes~\cite{volovikgorkov:1, volovikgorkov:2}. Nevertheless, an intriguing alternative, involving gapless BdG quasiparticles existing over an extended regime over the Fermi surface, is also possible. It gives rise to the notion of inflated nodes and Majorana-Fermi surfaces~\cite{yangsondhi, wilczek:1, wilczek:2, agterberg:1, agterberg:2, hirschfeld, herbut:1, herbut:2}.

Here we show that a spin- and valley-polarized quarter-metal (QM) phase in multilayer graphene heterostructures host a conducive podium where a marriage between PDW and isolated Fermi rings of normal fermions can take place due to the trigonal warping, resulting from the underlying three-fold rotational symmetry ($C_3$) of the honeycomb lattice in the normal state. Our findings are particularly pertinent in Bernal bilayer graphene (BBLG) and rhombohedral trilayer graphene (RTLG), where such a QM has been observed experimentally~\cite{BBLG:Exp1, BBLG:Exp2, BBLG:Exp3, RTLG:Exp1, RTLG:Exp2}, although superconductivity therein is yet to be realized. Nonetheless, our analysis can be extended and generalized to tetralayer, pentalayer, and hexalayer rhombohedral graphene to arrive at qualitatively similar conclusions where PDW phase has recently been observed in the close proximity to the QM~\cite{PDW:Exp1, PDW:Exp2, PDW:Exp3}. The current discussion and study are restricted to the superconducting instability of the QM phase in BBLG and RTLG. Notice that in rhombohedral stacking successive layers of honeycomb lattice shift by half of the lattice spacing. Consequently, the low-energy sites in BBLG [RTLG] form an effective honeycomb [prismlike] lattice, as shown in Fig.~\ref{fig:lattice}(a) [Fig.~\ref{fig:lattice}(b)]. As a result the Cooper pairs in the PDW phase, resulting from the pairing among fermions residing near one particular valley in QM phase, see Fig.~\ref{fig:lattice}(c), assumes Kekul\'e and columnar patterns in BBLG and RTLG, respectively, as shown in Figs.~\ref{fig:lattice}(d) and~\ref{fig:lattice}(e). Extending this observation we can conclude that two sets of low-energy sites in even and odd number of rhombohedral graphene multilayers constitute effective honeycomb and prismlike lattice, respectively, thereby fostering Kekul\'e and columnar PDWs therein in the proximity to the QM phase. Any quantitative discussion on such PDWs beyond BBLG and RTLG is, however, left for a future investigation.

The normal-state band structure in the QM featuring trigonally warped disconnected or annular or simply connected Fermi rings (depending on the chemical doping) and the resulting reconstructed band structure inside the PDW state showing isolated Fermi rings of normal fermions, connected by three-fold rotations, are displayed in Figs.~\ref{fig:SCbandBBLG} and~\ref{fig:SCbandRTLG} for BBLG and RTLG, respectively.~\footnote{Notice that in two and one dimensions, a Fermi surface becomes a Fermi ring and a pair of Fermi points, respectively.} This situation qualitatively mimics the \emph{two-fluid} scenario in He II~\cite{twofluid:1, twofluid:2, twofluid:3}. We begin the discussion by highlighting some essential background of these systems and subsequently presenting a brief summary of our main results.

\subsection{Background and main results}

Electronic bands in graphene heterostructures possess four-fold valley and spin degeneracy. On the other hand, experimentally observed global phase diagrams of BBLG and RTLG, subject to perpendicular electric displacement field ($D$), showcase electronic interaction driven cascade of spin and valley degeneracy lifting and the consequent formation of fractional metals. As such, both systems support (a) a nondegenerate (spin- and valley-polarized) QM at low doping, (b) a spin-polarized, but valley degenerate, half-metal at moderate doping, and (c) a spin and valley degenerate (fully unpolarized) metal at large doping~\cite{BBLG:Exp1, BBLG:Exp2, BBLG:Exp3, RTLG:Exp1, RTLG:Exp2}. Furthermore, in the proximity to the half-metal superconducting phases have been observed in these systems, with, however, an assistance from a weak in-plane magnetic field in BBLG~\cite{RTLG:Exp2, BBLG:Exp1}. Due to the spin-polarized nature of charged carriers in the half-metal, the paired state appearing in its close proximity can only be spin triplet in nature, devoid of the Pauli limiting in-plane magnetic fields.

\begin{figure*}[t!]
\centering
\includegraphics[width=0.90\linewidth]{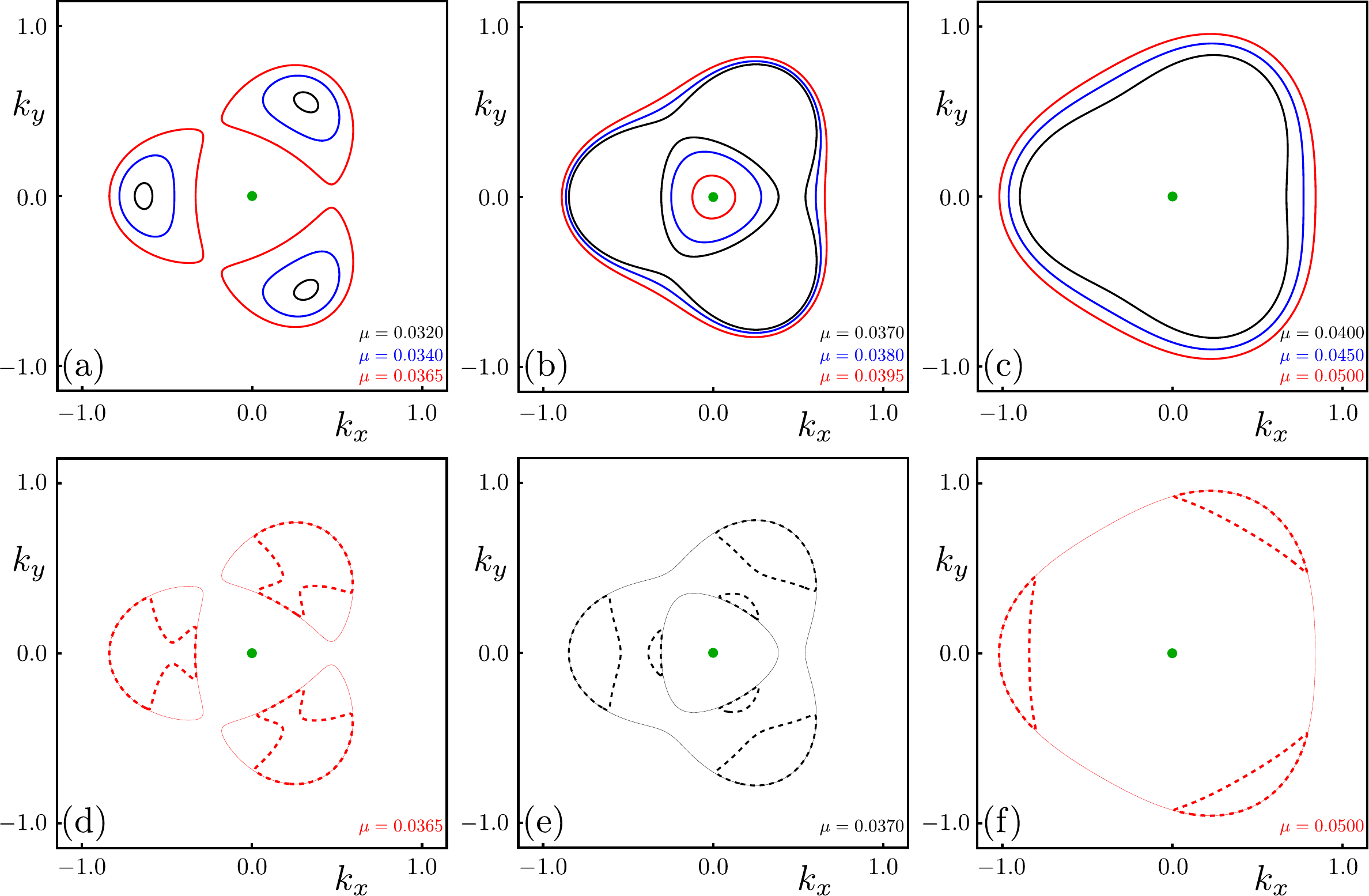}
\caption{Trigonally warped (a) disconnected, (b) annular, and (c) simply connected Fermi rings (FRs) in the quarter-metal phase of Bernal bilayer graphene, for small, moderate, and large chemical potential $\mu$, respectively, captured by constant energy contours (solid lines, color coded according to $\mu$ values, mentioned in legends). The momentum $\vec{k}$ (in units of $a^{-1}$, where $a$ is the lattice spacing) is measured from the valley momentum $\vec{K}$ (green dots) and $\mu$ (in units of the intralayer nearest-neighbor hopping amplitude $t$) is measured from the charge-neutrality point. For each FR topology we pick three representative values of $\mu$ to display their evolution with changing doping level. Isolated FRs are shown by the dashed lines in the presence of the Kekul\'e pair density wave [Fig.~\ref{fig:lattice}(d)] of amplitude $\Delta=2 \times 10^{-3}$ (in units of $t$) for (d) disconnected, (e) annular, and (f) simply connected normal-state FRs for one specific value of $\mu$ in each case (see legend) that result from the regions within the normal-state FRs, lacking the inversion symmetry under $\vec{k} \to -\vec{k}$. Rest of the parameter values are reported in Appendix~\ref{append:bandparameters}.       
}\label{fig:SCbandBBLG}
\end{figure*}

Surge of theoretical works~\cite{RTLGrecent:1, RTLGrecent:2, RTLGrecent:3, RTLGrecent:4, RTLGrecent:5, RTLGrecent:6, RTLGrecent:7, RTLGrecent:8, RTLGrecent:9, RTLGrecent:10, RTLGrecent:11, RTLGrecent:12, RTLGrecent:13, RTLGrecent:14, BBLGrecent:1, BBLGrecent:2, BBLGrecent:3} strongly suggests that the triplet paired state is possibly $f$ wave in nature, either mediated by repulsive Coulomb interactions in the spirit of the Kohn-Luttinger mechanism~\cite{RTLGrecent:3, RTLGrecent:4, RTLGrecent:5, RTLGrecent:6, RTLGrecent:7, RTLGrecent:8, RTLGrecent:9, BBLGrecent:2, BBLGrecent:3, kohn-Luttinger} or by optical phonons~\cite{RTLGrecent:1, BBLGrecent:1}. Although no superconductivity near the QM has been observed yet in BBLG and RTLG, its spin and valley polarization gives rise to fascinating possibilities. Firstly, when valley- and spin-polarized quasiparticles condense into Cooper pairs they give rise to a \emph{unique} superconducting ground state that assumes a spatially periodic structure of periodicity $2 \vec{K}$, where $\vec{K}$ is the valley momentum. It stands as an example of a PDW on crystalline honeycomb heterostructures. In the ordered phase, the PDW optimally gaps disjoint or annular or simply connected Fermi rings, leaving only isolated rings of normal fermions gapless, otherwise connected via threefold rotations. These outcomes for BBLG and RTLG are shown in Figs.~\ref{fig:SCbandBBLG} and~\ref{fig:SCbandRTLG}, respectively.

By projecting the local (momentum-independent) PDW onto the Fermi rings, we show that it always corresponds to an \emph{odd-parity} pairing, as it should be the case for spin- and valley-polarized systems to obey the Pauli exclusion principle. In BBLG, the odd-parity characteristic of the PDW is dominantly $p$-wave in nature, sourced by the trigonal warping term in the normal state that is captured by a linear in momentum Dirac Hamiltonian, also described by the odd-parity angular momentum $\ell=1$ or $p$ wave harmonics. By contrast, near the Fermi rings the PDW is dominantly $f$ wave in nature in RTLG, resulting from the angular momentum $\ell=3$ or $f$ wave harmonics describing the dominant cubic band dispersion in the normal state. Altogether, these findings manifest the paramount importance of the normal-state band structure in ensuring the requisite odd-parity nature of the PDW near the nondegenerate Fermi rings, realized either in the conduction or valence band.

Under the assumption that an effective pairing interaction ($g$) is operative only around the Fermi rings in the conduction or valence band, we self-consistently solve for the (dimensionless) amplitude of the PDW at zero temperature ($\Delta$) and its critical temperature ($t_c$) within a mean-field approximation. The solutions are shown in Fig.~\ref{fig:SolutionsBBLG} and Fig.~\ref{fig:SolutionsRTLG} for BBLG and RTLG, respectively. In brief, when the Fermi rings enjoy even a partial nesting under $\vec{k} \to -\vec{k}$ around the valley momentum $\vec{K}$, we find non-trivial solutions of $\Delta$ for arbitrarily weak effective attractive interaction, irrespective of the topology of the Fermi rings (disjoint or annular or simply connected). Otherwise, non-trivial solution of $\Delta$ can only be found beyond a critical interaction strength in the absence of such nesting, which, however, can only occur for small disjoint Fermi rings in the normal state [see Figs.~\ref{fig:SolutionsBBLG}(a)-~\ref{fig:SolutionsBBLG}(c) and~\ref{fig:SolutionsRTLG}(a)-{fig:SolutionsRTLG}(c)]. Due to the presence of isolated Fermi rings in the paired state, finite $t_c$ typically requires a threshold value of interaction strength, implying that the paired state can only exists at zero temperature for interaction strength below such threshold value [see Figs.~\ref{fig:SolutionsBBLG}(d)-~\ref{fig:SolutionsBBLG}(f) and~\ref{fig:SolutionsRTLG}(d)-~\ref{fig:SolutionsRTLG}(f). On the other hand, if the pairing interaction is assumed to be operative on the portion of the Fermi rings, enjoying the inversion symmetry $\vec{k} \to -\vec{k}$ around the valley momentum $\vec{K}$, then the scaling of $t_c$ with interaction strength closely follows that of $\Delta$ with $g$ for any choice of $\mu$, as shown in Figs.~\ref{fig:SolutionsBBLG}(d)-~\ref{fig:SolutionsBBLG}(f) and~\ref{fig:SolutionsRTLG}(d)-~\ref{fig:SolutionsRTLG}(f). It should, however, be noted that due to the reduced dimensionality of the system, the mean-field solutions of $t_c$ corresponds to a crossover temperature associated with the formation of the Cooper pairs. However, their coherent condensation at the macroscopic scale, yielding superconductivity in the system, occurs at a lower temperature through the Kosterlitz-Thouless transition, going beyond the scope of this study.

\subsection{Organization}

The rest of the paper is organized as follows. In the next section (Sec.~\ref{Sec:EffectiveHamiltonian}), we derive the effective Hamiltonian for the QM phase in BBLG and RTLG. In Section~\ref{Sec:PDW}, we introduce the only local pairing, namely the Kekul\'e (in BBLG) or columnar (in RTLG) PDW, allowed by symmetry in valley and spin polarized QM in these systems, and identify its odd-parity nature. Section~\ref{Sec:meanfield} is devoted to a mean-field analysis of such an odd-parity PDW to compute its amplitude at zero temperature and the transition temperature as functions of an effective attractive interaction with varying chemical potential in the system. A summary of our results and discussions on related topics are presented in Sec.~\ref{Sec:summary}. Additional details of this study are relegated to four appendixes. Namely, in Appendix~\ref{append:QMuniformmass}, we present a universal mechanism for the cascade of degeneracy lifting in rhombohedral graphene heterostructures, formation of half-metal and QM therein, and identification of the amplitude of the uniform gap in the QM phase. Various band parameters and strength of the $D$ field required for the QM phase are summarized in Appendix~\ref{append:bandparameters}. Band diagonalization procedure unfolding the odd-parity nature of the PDW state is detailed in Appendix~\ref{append:banddiaginalization}. Convergence of our numerical solutions for the self-consistent gap equation is discussed in Appendix~\ref{append:convergence}, and shown in Figs.~\ref{fig:ConvergenceBBLG} and~\ref{fig:ConvergenceRTLG}.

\section{QM: Effective Hamiltonian}~\label{Sec:EffectiveHamiltonian}

We begin the discussion by deriving the effective two-band Hamiltonian for the QM in BBLG and RTLG. We construct a minimal model involving intralayer nearest-neighbor ($t$) and direct interlayer ($t_\perp$) hopping. Then, two out of four and six bands in BBLG and RTLG, respectively, closest to the zero energy feature quadratic and cubic band touching degeneracy at two inequivalent valleys at momenta $\pm \vec{K}$. Here, $\vec{K}=2\pi (1,\sqrt{3})/(3a)$ and $a$ is the lattice spacing. The remaining, split-off bands reside at higher energies $t_\perp \approx 200$ meV (roughly)~\cite{graphene:RMP}. The fermion operators that encompass all these degrees of freedom can be organized in spinors, given by
\begin{equation*}
\Psi^\top(\vec{q})= \left\{
\begin{array}{rl}
& [c_{b_1},c_{a_2},c_{a_1},c_{b_2}](\vec{q}) \quad \text{for BBLG} \\
& {[c_{b_1},c_{a_3},c_{a_1},c_{b_2},c_{a_2},c_{b_3}](\vec{q})} \quad \text{for RTLG}
\end{array} \right.
.
\end{equation*}
Here $c_{sl}(\vec{q})$ is a fermion annihilation operator on sublattice $s=a,b$, layer $l=1,2$ (BBLG) or $l=1,2,3$ (RTLG), and with lattice momentum $\vec{q}$. The split-off bands dominantly reside on the pairs of overlapping sites, namely, $b_1$ and $a_2$ in BBLG [Fig.~\ref{fig:lattice}(a)], and $b_1$ and $a_2$, as well as $b_2$ and $a_3$ in RTLG [Fig.~\ref{fig:lattice}(b)]. The gapless bands live on the remaining two set of sites in both systems. The direct hopping with amplitude $t_3$ between them gives rise to trigonal warping. It splits the quadratic and cubic touching points into four and three Dirac points in a $C_3$ symmetric manner, respectively, thereby preserving local nodal topology near each valley. The effect of the displacement $D$ field is taken into account in terms of the on-site potential $V$. It appears with opposite signs on the top and bottom layers in BBLG and RTLG, while in the middle layer of RTLG $V=0$.

With the above considerations, the tight-binding Hamiltonian takes the generic block matrix form
\begin{align}
H=
\begin{pmatrix}
H_{\rm LL} & H_{\rm LH} \\
H_{\rm HL} & H_{\rm HH} \\
\end{pmatrix},
\end{align} 
where $H_{\rm LL}$ ($H_{\rm HH}$) is the Hamiltonian within the subspace of only the low- (high-) energy sites, while $H_{\rm LH}$ and $H_{\rm HL}$ capture the coupling between them with $H_{\rm HL}=H^\dag_{\rm LH}$. For BBLG these matrix components are
\begin{eqnarray}
H_{\rm LL} &=&
\begin{pmatrix}
V & t_3 \gamma \\
t_3 \gamma^\ast & -V 
\end{pmatrix}, \;\;
H_{\rm HH} =
\begin{pmatrix}
V & t_\perp \\
t_\perp & -V 
\end{pmatrix}, \nonumber \\
\text{and} \:\:
H_{\rm LH} &=& \begin{pmatrix}
t \gamma^\ast & 0 \\
0 & t \gamma
\end{pmatrix}, 
\end{eqnarray}
where
\begin{equation} 
\gamma \equiv \gamma(\vec{q})=\sum_{i=1}^3\exp [i \vec{q} \cdot \boldsymbol{\delta}_i], 
\end{equation}
and $\boldsymbol{\delta}_1=-a \left( 1,1/\sqrt{3}\right)/2$, $\boldsymbol{\delta}_2=a \left( 1,-1/\sqrt{3}\right)/2$, and $\boldsymbol{\delta}_3= a\left( 0, 1/\sqrt{3}\right)$ are three nearest-neighbor lattice vectors. On the other hand, for RTLG we have
\begin{align}
& H_{\rm LL} =
\begin{pmatrix}
V & t_3 \\
t_3 & -V 
\end{pmatrix}, \;\;
H_{\rm HH}=
\begin{pmatrix}
V & t_\perp & 0 & 0 \\
t_\perp & 0 & t \gamma^\ast & 0 \\
0 & t \gamma & 0 & t_\perp \\
0 & 0 & t_\perp & -V
\end{pmatrix}, \nonumber \\
& \text{and} \;\;
H_{\rm LH} =
\begin{pmatrix}
t \gamma^\ast & 0 & 0 & 0\\
0 & 0 & 0 & t \gamma 
\end{pmatrix}.
\end{align}
Notice that two low-energy triangular sublattices constitute a honeycomb lattice in BBLG and prismlike  lattice in RTLG, as shown in Figs.~\ref{fig:lattice}(a) and~\ref{fig:lattice}(b), respectively. Therefore, the trigonal warping (hopping proportional to $t_3$) is devoid of any momentum dependence in RTLG.

To capture the low-energy band structure of these systems, we integrate out the split-off bands or equivalently the high-energy sites.
The renormalized Hamiltonian describing an effective two-band model is given by $H_{\rm ren}=H_{\rm LL}-H_{\rm LH}\ H^{-1}_{\rm HH}\ H_{\rm HL}$~\cite{dantasroy2021, panigrahi2022}, leading to 
\begin{align}
H_{\rm ren}=
\begin{pmatrix}
m(\vec{q}) & \alpha^\ast(\vec{q}) \\
\alpha(\vec{q}) & -m(\vec{q})
\end{pmatrix}.
\end{align}
For BBLG, we find 
\allowdisplaybreaks[4]
\begin{align} 
m= V \Big[1-\frac{t^2 |\gamma|^2}{t_\perp^2+V^2}\Big]
\; \text{and} \; 
\alpha = t_3 \gamma -\frac{t^2 t_\perp \gamma^2}{t_\perp^2+V^2}, 
\end{align}
and for RTLG we obtain
\begin{align}~\label{eq:RTLGbeforeexpansion}
m = V \Big[1-\frac{t^4 |\gamma|^4}{t_\perp^4 + t^2 V^2 |\gamma|^2}\Big]
\; \text{and} \;
\alpha = t_3 + \frac{t^3 t_\perp^2 \gamma^3}{t_\perp^4 + t^2 V^2 |\gamma|^2}.
\end{align}
In the last two equations, we used $m=m(\vec{q})$ and $\alpha=\alpha(\vec{q})$ for brevity. The low-energy model near each valley can now be obtained by expanding $H_{\rm ren}$ around it, about which more in a moment. Due to the low atomic number of carbon, the spin-orbit coupling in graphene heterostructures is negligible, and inclusion of spin degrees of freedom leads to a mere doubling of the noninteracting Hamiltonian $H_{\rm ren}$.

\begin{figure*}[t!]
\includegraphics[width=0.90\linewidth]{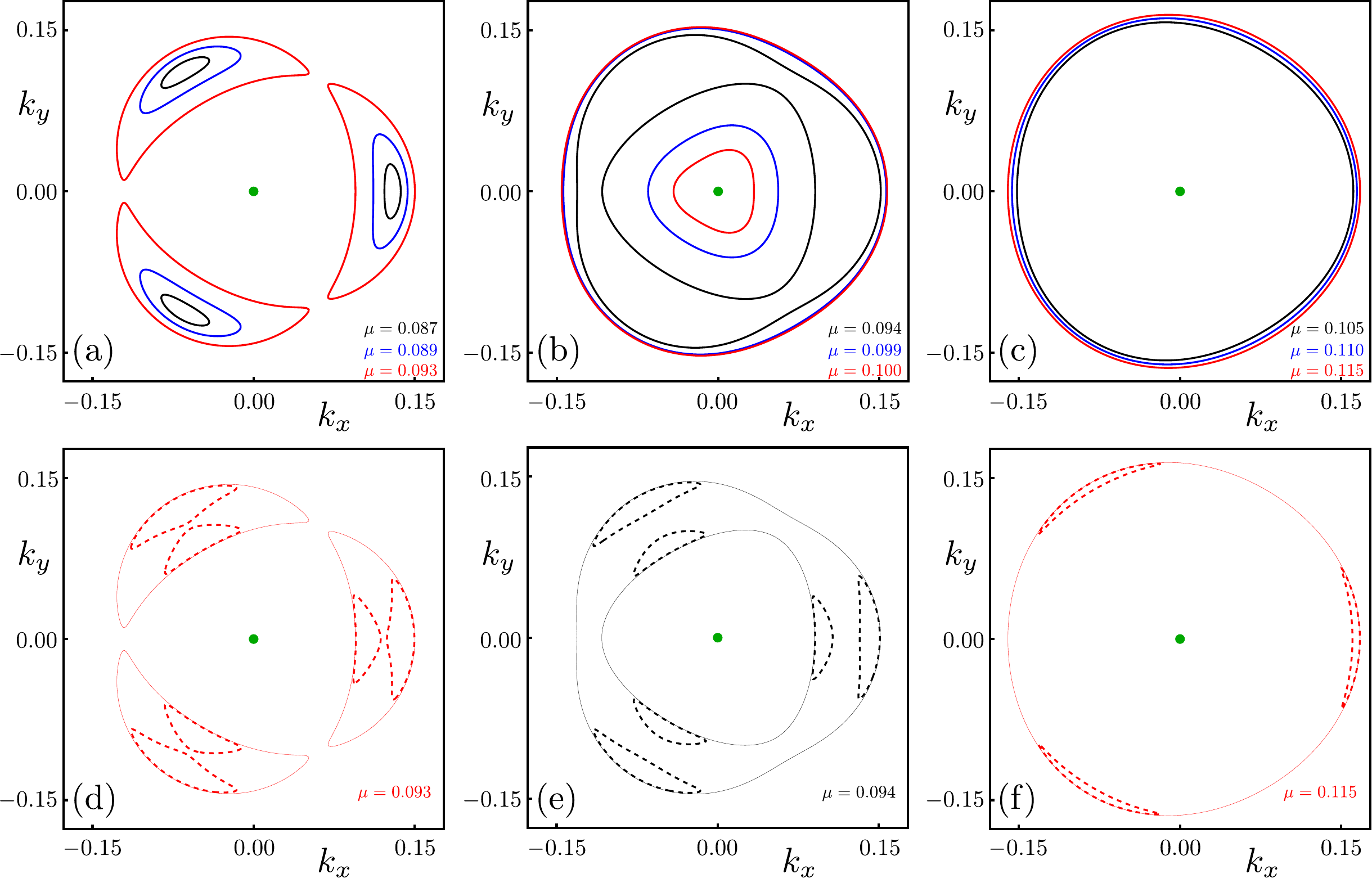}
\caption{Same as Fig.~\ref{fig:SCbandBBLG}, but for rhombohedral trilayer graphene. The values of the chemical potential $\mu$ are mentioned in the legend of each subfigure. The lower row is obtained in the presence of the columnar pair-density-wave [Fig.~\ref{fig:lattice}(e)] of amplitude $\Delta=2 \times 10^{-3}$. The rest of the parameter values are mentioned in Appendix~\ref{append:bandparameters}. 
}\label{fig:SCbandRTLG}
\end{figure*}

So far, we dealt with noninteracting fermions in BBLG and RTLG to arrive at their low-energy description. In order to construct an effective model for the valley- and spin-polarized QM, we need to consider the symmetry breaking in these systems. Fractional (half and quarter) metals stem from the simultaneous presence of multiple competing orders, for which the corresponding matrix operators \emph{commute} with each other~\cite{RTLGrecent:6, BBLGrecent:2}. The $D$-field induces a sublattice or equivalently a layer polarization of average electronic density is captured by $m(\vec{q})$. On the other hand, on-site Hubbard repulsion driven layer antiferromagnet ($\Delta_{\rm LAF}$) results in staggered pattern of electronic spin between the sublattices or layers~\cite{szaboroy:selectionrule}. The matrix operators for these two orders commute, and their simultaneous presence lifts the spin-degeneracy from electronic bands, leaving their valley degeneracy unaffected. A suitable choice of chemical doping then gives rise to a spin-polarized half-metal, detailed in Appendix~\ref{append:QMuniformmass}.

While on-site repulsion is the dominant short-range interaction in graphene-based systems~\cite{katsnelson:hubbard}, the next relevant component of Coulomb repulsion in BBLG and RTLG subject to the $D$ field is the intralayer next-nearest-neighbor repulsion. In a spin-polarized band structure, it induces quantum anomalous Hall order ($\Delta_{\rm QAH}$)~\cite{szaboroy:selectionrule, haldane}. This mass order produces a net Berry curvature by polarizing two valleys of opposite Berry phase, and features intra-sublattice circulating currents. Hence the corresponding matrix operator is diagonal in both valley and sublattice spaces, and commutes with those for layer polarized and layer antiferromagnet orders. Concomitantly, nucleation of the quantum anomalous Hall order in a spin-polarized half-metal lifts the remaining valley degeneracy, giving birth to a QM at low chemical doping, detailed in Appendix~\ref{append:QMuniformmass}. It should be noted that interlayer nearest-neighbor Coulomb repulsion results in a layer polarization of electronic density. This symmetry is, however, already broken by the $D$-field. Once the simultaneous presence of layer antiferromagnet and quantum anomalous Hall order is taken into account, $m(0) \equiv m_0=|V-\Delta_{\rm LAF}-\Delta_{\rm QAH}|<V$.

To arrive at the continuum description of QM in BBLG and RTLG, next we expand $\gamma(\vec{q}) \equiv \gamma(\vec{K}+\tilde{\vec{q}})$ to the \emph{linear} order in $\tilde{\vec{q}}$ around one of the valleys, otherwise picked spontaneously via symmetry breaking, which we choose to be at $+\vec{K}$ for concreteness. Subsequently, we rotate the momentum axis by $\pi/3$ about the $z$ direction taking $\tilde{\vec{q}} \to \vec{k}$, yielding $\gamma(\vec{k})=-\sqrt{3}(k_x+i k_y)/2$. After some algebraic simplifications, we then arrive at the effective single-particle Hamiltonian describing a QM in BBLG in the continuum limit 
\begin{eqnarray}~\label{eq:QMBBLGHamil}
H^{\rm BBLG}_{\rm QM} &=& \alpha_2 [\beta_1 d_1(\vec{k}) + \beta_2 d_2(\vec{k})] +
\alpha_1 \big[\beta_1 p_1(\vec{k}) \nonumber \\
&-& \beta_2 p_2(\vec{k})] 
+ \beta_3 \left[ m_0- \frac{\alpha_2}{t_\perp} V |\vec{k}|^2 \right] -\mu \beta_0, 
\end{eqnarray}  
where the Pauli matrices $\{ \beta_\nu \}$ with $\nu=0,\cdots, 3$ operate on the low-energy sites, $\mu$ is the chemical potential, measured from the charge neutrality point, $\vec{p}(\vec{k})=(k_x,k_y)$, $\vec{d}(\vec{k})=(k_x^2-k_y^2, 2k_x k_y)$, $\alpha_1= -\sqrt{3}t_3/2$, and $\alpha_2=- 3 t^2 t_\perp/(4\left[ t^2_\perp + V^2\right])$. The effective single-particle Hamiltonian describing a QM in RTLG is given by  
\begin{eqnarray}~\label{eq:QMRTLGHamil}
H^{\rm RTLG}_{\rm QM} &=& \beta_1 \left[ \alpha_3 f_1(\vec{k}) + t_3 \right] + \beta_2 \alpha_3 f_2(\vec{k}) \nonumber \\
&+& \beta_3 \left[ m_0- \frac{\alpha^4_0}{t^4_\perp} V |\vec{k}|^4 \right] -\mu \beta_0,
\end{eqnarray}  
where $\vec{f}=(k_x^3-3 k_x k_y^2, -k_y^3 + 3 k_x^2 k_y)$, $\alpha_0=\sqrt{3} t/2$, and $\alpha_3=-\left(\sqrt{3}t/2 \right)^3/t^2_\perp$. While arriving at the above form of $H^{\rm RTLG}_{\rm QM}$ we take $t_\perp^4 + t^2 V^2 |\gamma|^2 \to t_\perp^4$ in the denominator of the second terms of both $m(\vec{q})$ and $\alpha(\vec{q})$ in Eq.~\eqref{eq:RTLGbeforeexpansion}, as $|\gamma|^2 \to 0$ near the Dirac points or valleys.

A final comment in this context is due at this stage. Notice that the term proportional to $\beta_3$, namely $m (\vec{k})$ represents a momentum-dependent Semenoff-type mass, that opens a gap near the charge neutrality point~\cite{semenoff}. Due to the requisite simultaneous presence of layer polarization, layer antiferromagnet, and anomalous Hall orders to realize a nondegenerate QM, its momentum independent component is given by $m(0)=m_0=|V-\Delta_{\rm LAF}-\Delta_{\rm QAH}|$, as mentioned previously. By contrast, momentum dependence of $m(\vec{k})$ solely originates from the split-off band projection in the presence of the $D$-field. As a result, $m(\vec{k})$ changes sign at some momentum $|\vec{k}|$, thereby yielding disconnected and annular Fermi rings in the normal state for small and moderate chemical potential, respectively, finally giving way to simply connected Fermi rings at larger doping. Such changes in the Fermi ring topology take place through Lifshitz transitions. In the top rows of Figs.~\ref{fig:SCbandBBLG} and~\ref{fig:SCbandRTLG}, we depict the evolution of the Fermi rings in BBLG and RTLG, respectively. If, on the other hand, we introduce effects of displacement field after the band projection, then $m(\vec{k})$ becomes momentum independent. It then produces a uniform mass gap near the charge neutrality, resulting in only a simply connected regular Fermi ring upon doping. Next, we proceed to discuss the PDW instability of QM.

\section{Odd-parity pair-density-wave}~\label{Sec:PDW}

As a consequence of the QM being valley- and spin-polarized, the only available channel for electrons to condense into  Cooper pairs is intra-valley and spin-triplet in nature. Intra-valley Cooper pairs are formed by electrons carrying opposite momentum ($+\vec{k}$ and $-\vec{k}$) with respect to the valley momentum $\vec{K}$, resulting in a center of mass momentum $2 \vec{K}$. The \emph{unique} available pairing channel arising from the QM is therefore a commensurate FFLO or a PDW. To formally incorporate superconductivity, we introduce a Nambu basis as $\Psi^N_{\vec{k}}=[ \Psi_{\vec{k}}, \Psi^\ast_{-\vec{k}}]^\top$, where $\Psi_{\vec{k}}=[c_{\vec{k},a}, c_{\vec{k},b}]^\top$ is a two-component spinor describing electrons in the effective low-energy theory. Here, we suppress the redundant layer index. The corresponding BdG Hamiltonian reads as
\begin{align}~\label{eq:HNambu}
H^{j}_{\rm BdG}={\rm diag}.\;[H^{j}_{\rm QM}(\vec{k}),-\left(H^{j}_{\rm QM}(-\vec{k})\right)^\top] + H_{\rm pair},
\end{align}
for $j=$ BBLG and RTLG, where $\top$ corresponds to transposition. Here $H_{\rm pair}=\Delta(\eta_1 \cos\phi+\eta_2 \sin\phi) M$ is the pairing Hamiltonian with amplitude $\Delta$ and superconducting U(1) phase $\phi$. Pauli matrices $\{ \eta_i \}$ operate on the Nambu sector. Due to the particle-hole symmetry in the BdG formalism $M$ is constrained to be purely imaginary, leaving the \emph{unique} choice of pairing matrix $M=\beta_2$.~\footnote{The Nambu spinor is endowed with the antiunitary charge conjugation symmetry under $C=\eta_1 \mathcal{K}$, where $\mathcal{K}$ is complex conjugation. For $(\Psi^N_{\vec{k}})^\dag H_{\rm pair} \Psi^N_{\vec{k}}$ to be nonvanishing $H_{\rm pair}$ and $C$ must anticommute, constraining Hermitian $M$ to be purely imaginary. Since $M$ must be a two-dimensional purely imaginary Hermitian matrix that yields a unique choice of $M=\beta_2$.} Therefore, Cooper pairs are formed by combining electrons from two low-energy sublattices of BBLG and RTLG, residing on the complementary layers.

PDW ordering on honeycomb heterostructure results in a modulation of the superconducting amplitude with a periodicity of 2$\vec{K}$, which can only be accommodated within an enlarged unit cell containing six sites and nine bonds~\cite{roy-herbut:PDW, shi:PDWgraphene, sigrist:PDWgraphene}. As two low-energy sites $b_1$ and $a_2$ of BBLG constitute an emergent honeycomb lattice [see Fig.~\ref{fig:lattice}(a)], the PDW assumes the profile of a Kekul\'e pattern therein [see Fig.~\ref{fig:lattice}(d)], similar to the one in monolayer graphene~\cite{roy-herbut:PDW}. By contrast, two sets of low energy sites $b_1$ and $a_3$ residing on the bottom and top layers of RTLG, respectively, constitute an effective prismlike lattice [see Fig~\ref{fig:lattice}(b)]. As a consequence, RTLG harbors a \emph{columnar} PDW [see Fig~\ref{fig:lattice}(e)]. Nonetheless, in both systems the PDW preserves the 3-fold rotational symmetry. Notice that in contrast to pristine graphene, the PDW emerging from a spin and valley polarized QM of BBLG and RTLG loses the concept of valley exchange symmetry (even or odd)~\cite{roy-herbut:PDW}.

\begin{figure*}[t!]
\includegraphics[width=0.90\linewidth]{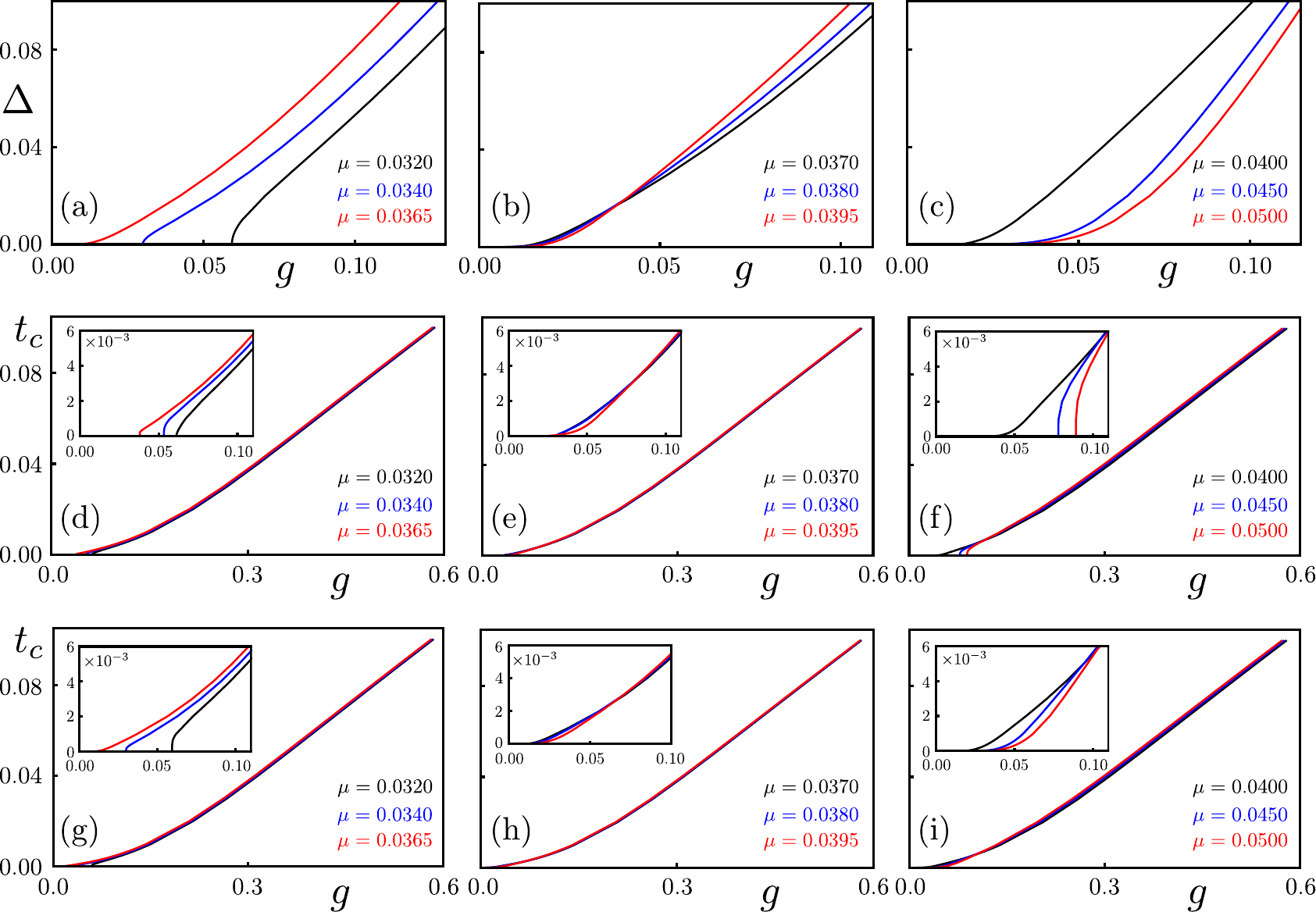}
\caption{Mean-field solutions (obtained numerically) of the amplitude ($\Delta$) of the Kekul\'e pair density wave at zero temperature as a function of the effective coupling constant ($g$) with an underlying (a) disjoint, (b) annular, and (c) simply connected Fermi rings [see Fig.~\ref{fig:SCbandBBLG}] in the normal state of the quarter-metal phase in Bernal bilayer graphene. For each Fermi ring geometry we choose three representative values of $\mu$ (mentioned in the legends), which are the same as in Fig.~\ref{fig:SCbandBBLG}. (d)-(f) Analogous to (a)-(c), respectively, but for the self-consistent solutions of the transition temperature ($t_c$), under the assumption that the pairing interaction is operative over the entire Fermi ring. (g)-(i) Analogous to (d)-(f), respectively, but obtained under the assumption that the pairing interaction is operative only over the portion of the Fermi rings enjoying the inversion symmetry under $\vec{k} \to -\vec{k}$ around the valley momentum $\vec{K}$. Here, $\mu$, $g$, $\Delta$, and $t_c$ are dimensionless and are measured in units of the intra-layer nearest-neighbor hopping amplitude of the honeycomb lattice ($t$). For technical details see Sec.~\ref{Sec:meanfield} and for discussions see Sec.~\ref{subsec:BBLGmeanfield}. Solutions of $\Delta$ and $t_c$ are obtained down to $10^{-6}$, and for their numerical convergence see Fig.~\ref{fig:ConvergenceBBLG}.  
}~\label{fig:SolutionsBBLG}
\end{figure*}

To gain insights into the symmetry of the PDW state, especially unfolding its odd-parity nature and the fingerprint of the normal state band structure, next we project the $H^{j}_{\rm BdG}$ on to the conduction band, fostering Fermi rings under the assumption that $\mu>0$. In this process, we only focus on the intra-band component of $H_{\rm pair}$ and neglect its inter-band component. This step is microscopically justified as the effective attractive or pairing interaction exists only near the Fermi ring realized on the conduction or valence band. The corresponding technical details are shown in Appendix~\ref{append:banddiaginalization}.

The reduced BCS Hamiltonian in BBLG reads 
\begin{eqnarray}~\label{eq:ReducedBCSBBLG}
H^{\rm BBLG}_{\rm BCS} &=& E_- \eta_0 + \left( E_+ -\mu \right) \eta_3 \nonumber \\
&+& \Delta \frac{\alpha_1}{\mu} \bigg( p_1(\vec{k})\eta_1 -p_2(\vec{k}) \eta_2 \bigg),
\end{eqnarray}
where we have retained the dominant and isotropic component of the pairing, the set of Pauli matrices $\{ \eta_\nu \}$ with $\nu=0,\cdots, 3$ acts on the particle-hole index, $E_\pm = (E_p \pm E_h)/2$ with 
\begin{eqnarray}~\label{eq:EphBBLG}
	E_{{\stackrel[h]{}{p}}}
=\sqrt{\alpha^2_2 |\vec{k}|^4 +\alpha^2_1 |\vec{k}|^2 \pm 2 \alpha_1 \alpha_2 |\vec{k}|^3 \cos(3 \phi_{\vec{k}})+ m^2(\vec{k})},
\nonumber \\
\end{eqnarray}
where $\phi_{\vec{k}}=\tan^{-1}(k_y/k_x)$ and $m(\vec{k})= m_0 - V \alpha_2 |\vec{k}|^2/t_\perp$. Some important comments are due at this stage, which qualitatively carry to RTLG, discussed shortly. Firstly, the term proportional to $\eta_0$ is nonvanishing only in the presence of trigonal warping in the system. This term is responsible for producing unpaired Fermi rings of normal fermions inside the paired state, which can be seen by diagonalizing $H^{\rm BBLG}_{\rm BCS}$, yielding the eigenvalue spectrum
\begin{equation}~\label{eq:EVBdGBBLG}
E^\tau_{\rm BdG}= E_- + \tau \sqrt{\left( E_+ -\mu \right)^2 + \Delta^2 \alpha^2_1 |\vec{k}|^2/\mu^2}
\end{equation}
for $\tau=\pm$. In particular, $E^+_{\rm BdG}$ reveals such Fermi rings (connected by three-fold rotations) as shown in the lower panel of Fig.~\ref{fig:SCbandBBLG}. Next, notice that the intra-band component of the PDW is \emph{odd} under the parity ($\vec{k} \to -\vec{k}$) and is proportional to the amplitude of the trigonal warping $\alpha_1$. This is so, because the trigonal warping is also expressed in terms of odd-parity functions $p_{1,2}(\vec{k})$, which takes the form of the Dirac Hamiltonian. As such, if we completely switch off the trigonal warping, then the intra-band component of the PDW is always identically \emph{zero}, since no other term in $H^{\rm BBLG}_{\rm QM}$ [see Eq.~\eqref{eq:QMBBLGHamil}] contains any function of momentum that is odd under the parity. For the exactness of this statement, see Appendix~\ref{append:banddiaginalization}. Since the bands in QM are completely non-degenerate they can only foster odd-parity pairing to fulfill the Pauli exclusion principle. In case of BBLG, the odd-parity pairing is sourced by the trigonal warping. Normal state band structure this way dictates the necessary symmetry of a microscopic local (momentum-independent) paired states near the Fermi ring (or surface)~\cite{banddiag:1, banddiag:2, banddiag:3}.

A similar exercise, also detailed in Appendix~\ref{append:banddiaginalization}, leads to the following reduced BCS Hamiltonian in the QM of RTLG in the presence of a columnar PDW
\allowdisplaybreaks[4] 
\begin{eqnarray}~\label{eq:ReducedBCSRTLG}
H^{\rm RTLG}_{\rm BCS} &=& E_- \eta_0 + \left( E_+ -\mu \right) \eta_3 \nonumber \\
&+& \Delta \frac{\alpha_3}{\mu} \bigg( f_1(\vec{k})\eta_1 -f_2(\vec{k}) \eta_2 \bigg),
\end{eqnarray}
where $E_\pm = (E_p \pm E_h)/2$ and now
\allowdisplaybreaks[4]
\begin{eqnarray}~\label{eq:EphRTLG}
	E_{{\stackrel[h]{}{p}}}
=\sqrt{\alpha^2_3 |\vec{k}|^6 +t^2_3 \pm 2 t_3 \alpha_3 |\vec{k}|^3 \cos(3 \phi_{\vec{k}})+ m^2(\vec{k})} \nonumber \\
\end{eqnarray}
and $m(\vec{k})=m_0-V \alpha^4_0 |\vec{k}|^4/t^4_\perp$. In this case, the odd-parity intra-band pairing is sourced by the component of the normal-state band dispersion that scales with the cubic power of momentum, captured by two planar $f$ wave harmonics that are also odd under the parity. Therefore, in RTLG even if we switch off the trigonal warping, the columnar PDW shown in Fig.~\ref{fig:lattice}(e), give rise to odd-parity $f$ wave pairing, which is dictated by the normal-state band geometry. The trigonal warping produces isolated Fermi rings that are connected by three-fold rotations, and are shown in the lower panel of Fig.~\ref{fig:SCbandRTLG} from the eigenvalues (namely $E^+_{\rm BdG}$) of $H^{\rm RTLG}_{\rm BCS}$, given by    
\begin{equation}~\label{eq:EVBdGRTLG}
E^\tau_{\rm BdG}= E_- + \tau \sqrt{\left( E_+ -\mu \right)^2 + \Delta^2 \alpha^2_3 |\vec{k}|^6/\mu^2}
\end{equation}
for $\tau=\pm$. Next, we proceed to compute the pairing amplitude ($\Delta$) and the associated transition temperature ($T_c$) within a mean-field approximation in BBLG and RTLG starting from the corresponding reduced BCS Hamiltonian shown in Eqs.~\eqref{eq:ReducedBCSBBLG} and~\eqref{eq:ReducedBCSRTLG}, respectively.

\section{Mean-field analysis}~\label{Sec:meanfield}

At first we assume that the effective attractive interaction ($g$) giving rise to the PDW in the QM phase in BBLG and RTLG is insensitive to the underlying Fermi ring geometry. Under this assumption, the free-energy density ($F$) within the mean-field approximation at temperature $T$ reads as 
\begin{equation}~\label{eq:freeenergytotal}
F=\frac{\Delta^2}{2g} - k_B T \sum_{\tau=\pm} \int \frac{d^2 \vec{k}}{(2\pi)^2} \ln \left[\cosh \bigg( \frac{|E^\tau_{\rm BdG}(\vec{k})|}{2 k_B T} \bigg)\right],
\end{equation}
where $k_B$ is the Boltzmann constant. The expressions for $E^\tau_{\rm BdG}(\vec{k})\equiv E^\tau_{\rm BdG}$ are given in Eqs.~\eqref{eq:EVBdGBBLG} and~\eqref{eq:EVBdGRTLG} for BBLG and RTLG, respectively. Minimizing $F$ with respect to the pairing amplitude $\Delta$, we arrive at the self-consistent mean-field gap equation 
\begin{equation}~\label{eq:gapequationfull}
\frac{1}{g}= \sum_{\tau=\pm} \int \frac{d^2 \vec{k}}{(2\pi)^2} \tanh\left( \frac{|E^\tau_{\rm BdG}(\vec{k})|}{2 k_B T} \right) \frac{\partial |E^\tau_{\rm BdG}(\vec{k})|}{\partial \Delta^2}.
\end{equation}
which we solve in two extreme limits. (1) At $T=0$, where we seek to obtain $\Delta$ as a function of $g$ from 
\begin{equation}~\label{eq:gapzerotemp}
\frac{1}{g}= \sum_{\tau=\pm} \int \frac{d^2 \vec{k}}{(2\pi)^2} \frac{\partial |E^\tau_{\rm BdG}(\vec{k})|}{\partial \Delta^2},
\end{equation}  
and (2) near the transition temperature $T_c$, which can be obtained by taking the derivative on the right hand side of Eq.~\eqref{eq:gapequationfull} in the limit $\Delta \to 0$ and setting $\Delta=0$ in tangent hyperbolic functions. In what follows, we set $k_B=1$, and measure $g$, $\Delta$, and $T$ in units of $t$, which then become dimensionless, and we denote the dimensionless transition temperature by $t_c$.

Alternatively, we can assume that the pairing interaction is operative only on the portions of the Fermi rings that enjoy nesting under $\vec{k} \to -\vec{k}$ since only such segments of the Fermi ring can be gapped out by the pairing. Then the free energy density reads as 
\begin{equation}~\label{eq:freeenergypartial}
F=\frac{\Delta^2}{2 g}- 2 k_B T \int \frac{d^2 \vec{k}}{(2\pi)^2} 
\ln \left[ \cosh \bigg( \frac{E_{\rm BdG}(\vec{k})}{2 k_B T} \bigg) \right],
\end{equation}
where 
\begin{equation*}
E_{\rm BdG}(\vec{k})= \left\{
\begin{array}{rl}
\sqrt{\left( E_+ -\mu \right)^2 + \Delta^2 \alpha^2_1 |\vec{k}|^2/\mu^2} & \text{for BBLG} \\
\sqrt{\left( E_+ -\mu \right)^2 + \Delta^2 \alpha^2_3 |\vec{k}|^6/\mu^2} & \text{for RTLG}
\end{array} \right.
.
\end{equation*}
The self-consistent equation for the zero temperature pairing amplitude remains unchanged from Eq.~\eqref{eq:gapzerotemp}, as $E_-$ is independent of $\Delta$. But, the equation for the transition temperature modifies to 
\begin{equation}
\frac{1}{g}=2 \int \frac{d^2 \vec{k}}{(2\pi)^2} \left[ \tanh\left( \frac{|E_{\rm BdG}(\vec{k})|}{2 k_B T_c} \right) \frac{\partial |E_{\rm BdG}(\vec{k})|}{\partial \Delta^2} \right]_{\Delta \to 0}. 
\end{equation}
In both cases, we compute the integration over the momentum by replacing it with a summation over a range of momentum $-k_\star \leq k_{x,y} \leq k_\star$ and dividing the corresponding two-dimensional reciprocal space into a large number of grid points ($N$), such that the numerical summation always converges for sufficiently large $N$, which we discuss in Appendix~\ref{append:convergence}. Guided by the topology of the Fermi rings in the QM phase which includes disconnected, annular, and simply connected Fermi rings (see Figs.~\ref{fig:SCbandBBLG} and~\ref{fig:SCbandRTLG}), we choose $k_\star=0.20$ for BBLG and $k_\star=0.25$ in RTLG. This assumption is consistent with the BCS approximation, which assumes that the pairing interaction exists only in the close proximity to the Fermi ring or surface only around which sharp quasiparticle excitations reside~\cite{schieffer:book, tinkham:book, BCS:original}. Notice that different choices of $k_\star$ do not alter the outcomes \emph{qualitatively}, which we discuss next for BBLG and RTLG in two subsequent subsections.

\begin{figure*}[t!]
\includegraphics[width=0.90\linewidth]{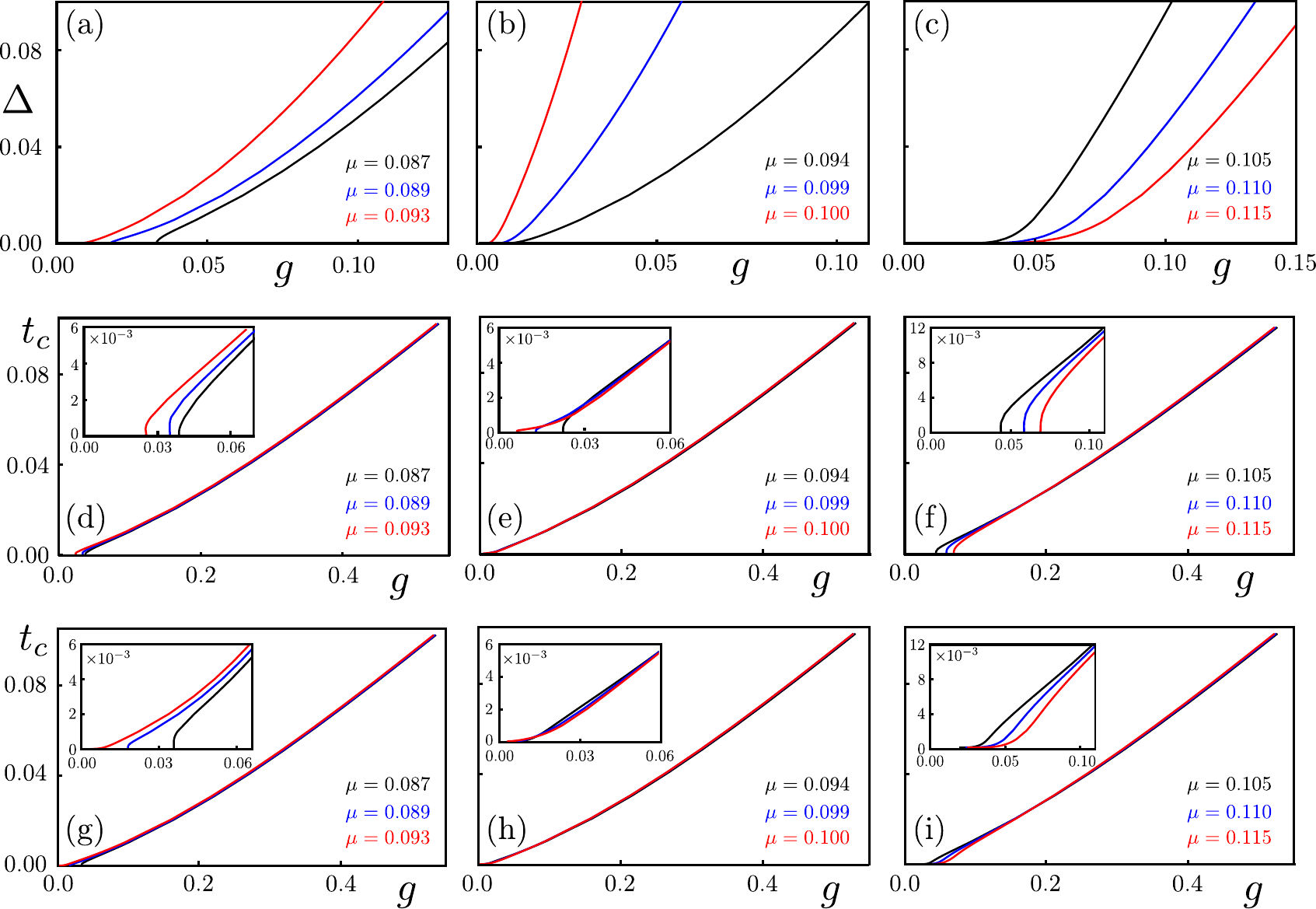}
\caption{Same as Fig.~\ref{fig:SolutionsBBLG}, but for the columnar pair density wave in rhombohedral trilayer graphene. For technical details see Sec.~\ref{Sec:meanfield} and for discussions see Sec.~\ref{subsec:BBLGmeanfield}. Here as well the solutions of dimensionless $\Delta$ and $t_c$ are obtained down to $10^{-6}$ and for the numerical convergence of these solutions, see Fig.~\ref{fig:ConvergenceRTLG}.  
}~\label{fig:SolutionsRTLG}
\end{figure*}

\subsection{Bernal bilayer graphene}~\label{subsec:BBLGmeanfield}

We first discuss the scaling of the pairing amplitude at zero temperature. When disjoint Fermi rings are small, devoid of any nesting under $\vec{k} \to -\vec{k}$ as is the case for sufficiently small $\mu$, the interaction strength needs to exceed a critical one for nontrivial solutions of $\Delta$. With increasing $\mu$, as soon as some segments of Fermi rings are nested, nontrivial solution of $\Delta$ appears even for infinitesimal $g$. These solutions are shown in Fig.~\ref{fig:SolutionsBBLG}(a). Otherwise, for a fixed $g$, $\Delta$ gets larger with increasing $\mu$ as the disjoint Fermi rings then enjoy larger degree of $\vec{k} \to -\vec{k}$ nesting. Next, we discuss the solutions of $\Delta$ with $g$ when the Fermi rings are simply connected, shown in Fig.~\ref{fig:SolutionsBBLG}(c). Although, in this case nontrivial $\Delta$ develops for infinitesimal $g$, their solutions are peculiar in the sense that for a fixed $g$, $\Delta$ gets smaller with increasing $\mu$. This observation can be justified in the following way. Notice that the requisite odd-parity nature of the paired state in BBLG is solely sourced by the trigonal warping. With increasing $\mu$, simply connected Fermi rings appear at progressively larger momentum, where the effects of trigonal warping gradually decrease, yielding a smaller gain in the condensation energy, thereby leading to the observed scaling behavior of $\Delta$ with varying $\mu$ for fixed $g$ values. In the intermediate doping regime, where the Fermi rings are annular, the solutions of $\Delta$ vs $g$ with varying $\mu$ display a crossover behavior between the ones we discussed above for disjoint and simply connected Fermi rings. In particular, for weaker (stronger) $g$, the scaling of $\Delta$ with varying $\mu$ mimics the one we observed for simply connected (disjoint) Fermi rings, as shown in Fig.~\ref{fig:SolutionsBBLG}(b), since for weaker (stronger) $g$, the density of states near the Fermi ring (degree of nesting among the Fermi rings) determines the scaling of $\Delta$ with $g$.

Scaling of the transition temperature ($t_c$) with varying $\mu$ and $g$ qualitatively follows the ones for the zero-temperature pairing amplitude. However, an important comment is due at this stage. If we assume that the pairing interaction is operative over the entire Fermi ring for any $\mu$, leading to the free-energy expression in Eq.~\eqref{eq:freeenergytotal}, often we find that finite $t_c$ can only be realized beyond critical interaction strength, which results from the existence of gapless normal fermions along the unnested segments of the Fermi rings that destroy superconducting order at any finite temperature due to fluctuations. These solutions are shown in Figs.~\ref{fig:SolutionsBBLG}(d), \ref{fig:SolutionsBBLG}(e), and~\ref{fig:SolutionsBBLG}(f) for disjoint, annular, and simply connected Fermi rings, respectively. By contrast, under the assumption that pairing interaction acts only on the portions of the Fermi rings enjoying nesting under $\vec{k} \to -\vec{k}$, leading to the free-energy shown in Eq.~\eqref{eq:freeenergypartial}, the scaling of $t_c$ with $g$ for various $\mu$ is \emph{qualitatively} identical to those for the zero-temperature pairing amplitude. For disjoint, annular, and simply connected Fermi rings the corresponding solutions for $t_c$ are shown in Figs.~\ref{fig:SolutionsBBLG}(g), \ref{fig:SolutionsBBLG}(h), and~\ref{fig:SolutionsBBLG}(i), respectively.

\subsection{Rhombohedral trilayer graphene}~\label{subsec:RTLGmeanfield}

In RTLG, the scaling of the pairing amplitude at zero temperature when the underlying Fermi rings are disjoint, shown in Fig.~\ref{fig:SolutionsRTLG}(a), is qualitatively similar to the ones we previously discussed for BBLG. Namely, with increasing $\mu$ as larger portion of the Fermi rings get nested we find amplification of $\Delta$. Although, the dependence of $\Delta$ on $g$ with varying $\mu$ with underlying simply connected Fermi rings in RTLG, shown in Fig.~\ref{fig:SolutionsRTLG}(c), is also similar to the ones for BBLG, the microscopic origin of such behavior in the former system is completely different. Notice that with increasing $\mu$ the simply connected Fermi rings are realized at larger momentum or energy. The dominant dispersion in this system that scales with the cubic power of momentum which also sources the odd-parity $f$-wave nature of the paired state, yields density of states (DOS) that scales as $\rho(E) \sim |E|^{-1/3}$. Consequently, with increasing $\mu$ the DOS near the Fermi energy ($\mu$) decreases as $|\mu|^{-1/3}$, leading to the suppression of $\Delta$ with increasing $\mu$ for any fixed $g$.

The parameter regime over which the QM in RTLG fosters annular Fermi rings is too narrow for us to identify a window of $\mu$ over which a crossover behavior (if it exists) between the ones shown in Figs.~\ref{fig:SolutionsRTLG}(a) and~\ref{fig:SolutionsRTLG}(c) can be seen clearly. Rather, in this regime we observe that with increasing $\mu$, the zero-temperature pairing amplitude increases for any fixed $g$ [see Fig.~\ref{fig:SolutionsRTLG}(b)]. This observation can be justified in the following way. When the system supports annular Fermi rings, with increasing $\mu$ the radius of inner Fermi ring shrinks faster than the growth of the radius of the outer ring. As the DOS increases (decreases) with decreasing (increasing) energy or momentum, the contribution from the inner ring is larger than that for the outer ring, leading to enhanced gain in condensation energy with increasing $\mu$, therefore yielding the observed scaling behavior. Otherwise, nontrivial solutions of $\Delta$ in this regime are found for arbitrarily weak interaction strength, as some portions of the Fermi rings are always nested.

Scaling of the transition temperature ($t_c$) in these three regimes in RTLG, shown in Figs.~\ref{fig:SolutionsRTLG}(d), \ref{fig:SolutionsRTLG}(e), and~\ref{fig:SolutionsRTLG}(f), are similar to the ones we previously discussed for BBLG. As these results are obtained by assuming that the pairing interaction is operative among all the quasiparticles which includes the ones that remain unpaired due to lack of $\vec{k} \to -\vec{k}$ nesting, often we find the necessity of a critical interaction strength only beyond which $t_c$ becomes finite. By contrast, if we assume that the attractive interaction is operative only among the nested portions of the Fermi rings then irrespective of its topology the scaling of $t_c$ with $g$ for varying $\mu$ is qualitatively identical to those for $\Delta$, as shown in Figs.~\ref{fig:SolutionsRTLG}(g), \ref{fig:SolutionsRTLG}(h), and~\ref{fig:SolutionsRTLG}(i).

\section{Summary and discussions}~\label{Sec:summary}

From a unified description of the spin and valley nondegenerate QM in BBLG and RTLG, featuring trigonally warped disconnected or annular or simply connected Fermi rings, here we identify a \emph{unique} superconducting instability in these systems. The paired state represents a PDW in both systems and supports isolated Fermi rings of normal fermions, connected via threefold rotations. Constant density of states produced by such Fermi rings should manifest in power-law scaling of the specific heat $C_v \sim T$, while the ratios of the longitudinal thermal conductivity ($\kappa_{jj}$) and inverse of nuclear magnetic relaxation time $1/(T_1)$ to temperature ($T$) approach constant values as $T \to 0$~\cite{roy-ghorashi-foster-nevidomskyy, carstentimm}. In this work, we also present the scaling of the amplitude of such a paired state at zero temperature and the associated transition temperature in BBLG and RTLG within a mean-field approximation, assuming that the effective attractive pairing interaction exists only within one band (assumed to be the conduction band for concreteness) fostering Fermi rings in the normal state and within a range of momentum around the Fermi rings. Most importantly, around the Fermi rings the PDW shows the requisite odd-parity nature, taking the form of a $p+ip$ pairing in BBLG sourced by the trigonal warping and $f+if$ pairing in RTLG stemming from the cubic band dispersion in the normal state. These outcomes altogether highlight the paramount importance of the normal state band dispersion in determining the requisite symmetry of the local PDW.

The PDW respectively assumes Kekul\'e and columnar shapes in BBLG and RTLG. Experimental observation of such novel quantum phases, however, faces a practical challenge as they require condensation of a macroscopic number of \emph{interlayer} Cooper pairs, which can considerably lower the critical temperature due to large interlayer distance $d \approx 3.4 \; \mathring{{\rm A}}$ in BBLG and $2d \approx 6.8 \; \mathring{{\rm A}}$ in RTLG. Nonetheless, application of external pressure in the stacking direction can considerably lower $d$ and that way favor the formation of the interlayer PDW. We note, however, PDW has recently been observed in experiments in the vicinity of the QM phase in tetralayer, pentalayer, and hexalayer rhombohedral graphene~\cite{PDW:Exp1, PDW:Exp2, PDW:Exp3}, making them promising to be observed in the proximity to the QM in BBLG and RTLG as well. Our theoretical predictions should therefore stimulate future experiments in these two systems. The proposed PDW, can, in principle, result from longer-ranged Hubbard-type repulsions, following the spirit of the Kohn-Luttinger mechanism~\cite{kohn-Luttinger}, which we leave for a future investigation.

\begin{figure*}[t!]
\includegraphics[width=0.90\linewidth]{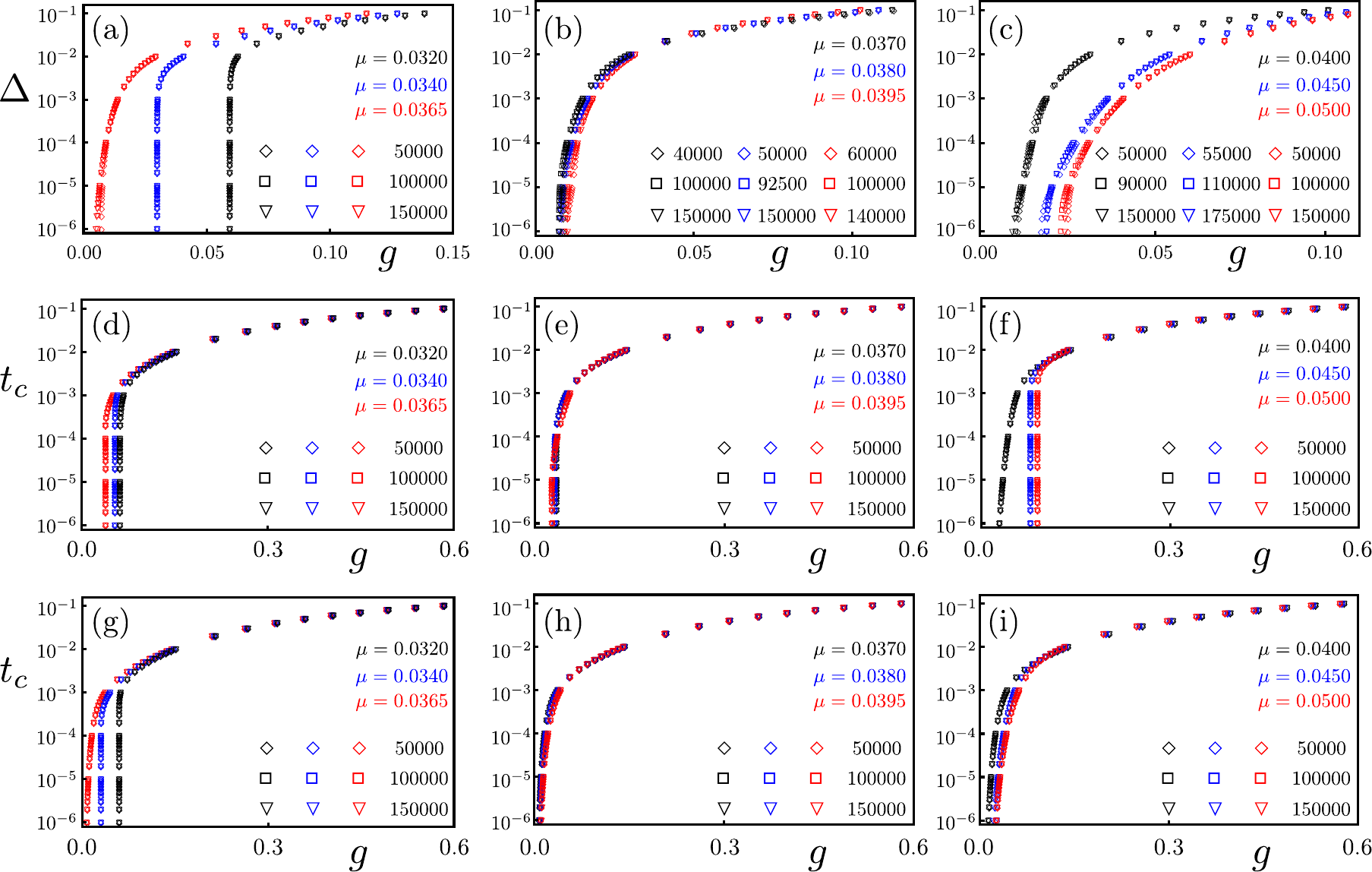}
\caption{Convergence of numerically obtained solutions of the pairing amplitude $\Delta$ at zero temperature shown in Figs.~\ref{fig:SolutionsBBLG}(a)-\ref{fig:SolutionsBBLG}(c) are shown here in (a)-(c), respectively. Convergence of the numerically obtained solutions of the transition temperature $t_c$ shown in Figs.~\ref{fig:SolutionsBBLG}(d)-(f) are shown here in (d)-(f), respectively. Convergence of the numerically obtained solutions of the transition temperature $t_c$ shown in Figs.~\ref{fig:SolutionsBBLG}(g)-(i) are shown here in (g)-(i), respectively. The numbers in different shapes (color coded according to $\mu$ values) correspond to the number of grid points ($N$) we divide the momentum space in the $(k_x, k_y)$ plane within the range $-0.2 \leq k_j \leq 0.2$ for $j=x$ and $y$ into after converting the integral over $\vec{k}$ in the gap equation into a discrete summation.  
}~\label{fig:ConvergenceBBLG}
\end{figure*}

We note that the nature of the pairing interaction, i.e. whether it depends on the shape of the Fermi ring or not, affects the scaling of the transition temperature ($t_c$) of the PDW with the effective attractive coupling constant ($g$). In Figs.~\ref{fig:SolutionsBBLG} (for BBLG) and~\ref{fig:SolutionsRTLG} (for RTLG), we have shown the dependence of $t_c$ on $g$ by assuming that $g$ is operative over the entire Fermi ring and thus being dependent on its shape (middle row), as well as under the assumption that $g$ is solely operative on the nested parts of the Fermi ring and thus being insensitive to its shape (bottom row). However, it is beyond the capacity of the mean-field theory to identify which situation can occur in reality. Notice that an effective attractive interaction responsible for pairing can result from purely repulsive electron-electron interactions following the general spirit of the Kohn-Luttinger mechanism~\cite{kohn-Luttinger}, which can be captured from a renormalization group calculation. Furthermore, the renormalization group approach is capable of accounting for the feedback from the underlying Fermi ring on the actual profile of the pairing interaction in an unbiased fashion which leaves its signature on the scaling of the transition temperature. We leave this exercise for a future investigation to unfold the interplay between the Fermi ring topology and effective attractive interaction yielding the PDW in a QM.

The requisite parent QM phase (normal state) and the resulting PDW therein can also be observed on optical lattices, where monolayer and crystalline multi-layer graphene have been realized~\cite{esslinger1}. Furthermore, in this setup density imbalance between the layers can be created (yielding a layer polarization), Hubbard repulsion driven anti-ferromagnetic order has already been observed~\cite{esslinger1}, and quantum anomalous Hall order has been engineered~\cite{esslinger2}. Thus, with these three necessary ingredients~\cite{RTLGrecent:6, BBLGrecent:2}, one can tune multi-layer optical honeycomb lattice into a QM phase, where the predicted PDW can be observed with tunable transition temperature. Although not experimentally pertinent, the Kekul\'e PDW can also condense from a QM in monolayer graphene, where it, however, represents a fully gapped state as the system is devoid of any trigonal warping~\cite{roy-herbut:PDW}. Altogether, the observed superconductivity, proximal to correlated ground states in crystalline~\cite{BBLG:Exp1, BBLG:Exp2, BBLG:Exp3, RTLG:Exp1, RTLG:Exp2, RHgraphite:Exp} and moir\'e graphene heterostructures~\cite{TBLG:Exp1, TBLG:Exp2, TBLG:Exp3, TBLG:Exp4}, along with the genuine possibility of electronic PDWs in BBLG and RTLG in the low doping regime originating from the QM place us at the dawn of the carbon age of superconductivity.

\acknowledgments

This work was supported by NSF CAREER Grant No.\ DMR-2238679 of B.R.\ (S.A.M.\ and B.R.) and Dr.\ Hyo Sang Lee Graduate Fellowship from Lehigh University (S.A.M.). B.R.\ thanks Christopher A.\ Leong and Vladimir Juri\v ci\' c for critical reading of the manuscript. Contributions of Andr$\acute{\mbox{a}}$s L. Szab$\acute{\mbox{o}}$ in the early stage of this project are acknowledged. 

\appendix

\section{Amplitude of uniform mass in QM}~\label{append:QMuniformmass}

In this appendix, we show the detailed derivation of the constant mass term, determined by $m(\vec{k}=0)$ entering in the effective single-particle Hamiltonian for the QM in BBLG and RTLG. For this purpose, we first write down the single-particle Hamiltonian in these two systems subject to perpendicular electric displacement field by invoking the valley and spin degrees of freedom. It can be readily achieved by extending the discussion from Sec.~\ref{Sec:EffectiveHamiltonian} upon noting the following. Due to the low atomic number of the constituting carbon atoms, the spin-orbit coupling in graphene-based heterostructures is negligibly small, and inclusion of the spin degrees of freedom leads to a mere doubling of the Hamiltonian that encompasses valley and sublattice or layer degrees of freedom. In Sec.~\ref{Sec:EffectiveHamiltonian}, we have shown the derivation of the low-energy Hamiltonian near the valley at $+\vec{K}$, from which we obtain that near the valley $-\vec{K}$ by noting the fact that under the exchange of two valleys $(k_x,k_y) \to (-k_x,k_y)$~\cite{HJR:2009PRB}. Thus, under $+\vec{K} \to -\vec{K}$ we have $(p_1,p_2) \to (-p_1,p_2)$, $(d_1,d_2) \to (d_1,-d_2)$, and $(f_1,f_2) \to (-f_1,f_2)$, and we then arrive at the following Hamiltonian 
\begin{eqnarray}~\label{eq:BBLGHamilTotal}
H_{\rm BBLG} &=& \alpha_2 [\Gamma_{001} d_1(\vec{k}) + \Gamma_{032} d_2(\vec{k})] +
\alpha_1 \big[\Gamma_{031} p_1(\vec{k}) \nonumber \\
&-& \Gamma_{002} p_2(\vec{k})] + \Gamma_{003} \left[ V- \frac{\alpha_2}{t_\perp} V |\vec{k}|^2 \right] -\mu \Gamma_{000} \nonumber \\ 
\end{eqnarray}  
for BBLG, while for RTLG we obtain
\begin{eqnarray}~\label{eq:RTLGHamilTotal}
H_{\rm RTLG} &=& \left[ \alpha_3 f_1(\vec{k}) \Gamma_{031} + \Gamma_{001} t_3 \right] + \Gamma_{002} \alpha_3 f_2(\vec{k}) \nonumber \\
&+& \Gamma_{003} \left[ V- \frac{\alpha^4_0}{t^4_\perp} V |\vec{k}|^4 \right] -\mu \Gamma_{000}.
\end{eqnarray}
Here, we have introduced a set of eight-dimensional Hermitian $\Gamma$ matrices, defined as $\Gamma_{\mu \nu \lambda}=\sigma_\mu \otimes \tau_\nu \otimes \beta_\lambda$, where $\otimes$ denotes the Kronecker product, and the sets of Pauli matrices $\{ \sigma_\lambda \}$, $\{ \tau_\lambda \}$, and $\{ \beta_\lambda \}$ act on spin, valley, and sublattice or layer degrees of freedom, respectively. As the layer and sublattice indices are locked in the low-energy theory for BBLG and RTLG, we define the corresponding eight-component spinor solely in terms of the sublattice (by omitting the layer index), valley, and spin degrees of freedom according to 
\begin{eqnarray}
\Psi^\top= [c_{a\uparrow}^{+\vec{K}},c_{b\uparrow}^{+\vec{K}},c_{a\uparrow}^{-\vec{K}},c_{b\uparrow}^{-\vec{K}}, c_{a\downarrow}^{+\vec{K}},c_{b\downarrow}^{+\vec{K}},c_{a\downarrow}^{-\vec{K}},c_{b\downarrow}^{-\vec{K}}]. \nonumber \\
\end{eqnarray}
The above two Hamiltonians possess four-fold degeneracy, resulting from the valley and spin degrees of freedom.

\begin{figure*}[t!]
\includegraphics[width=0.90\linewidth]{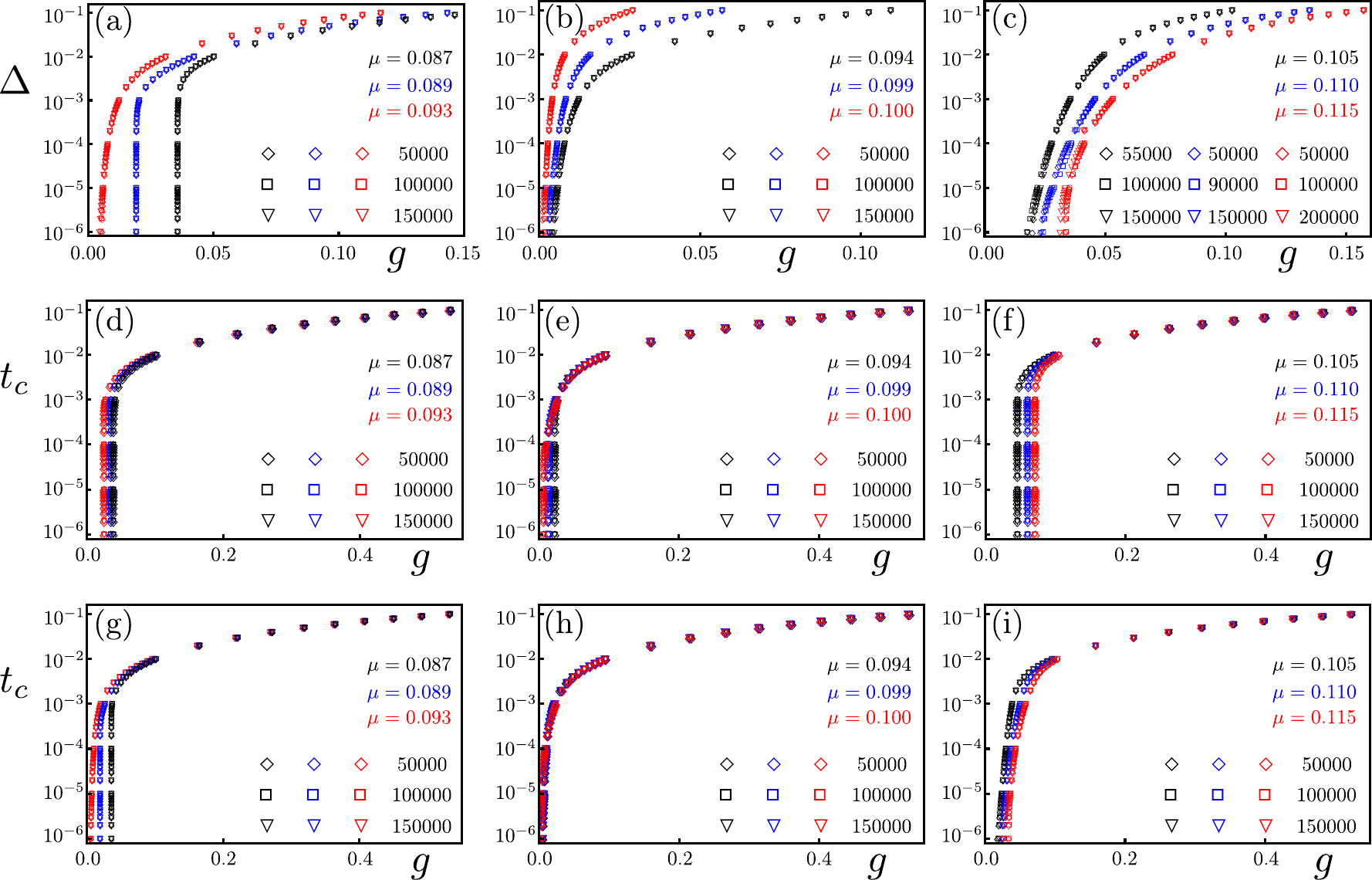}
\caption{Convergence of numerically obtained solutions of the pairing amplitude $\Delta$ at zero temperature shown in Figs.~\ref{fig:SolutionsRTLG}(a)-\ref{fig:SolutionsRTLG}(c) are shown here in (a)-(c), respectively. Convergence of the numerically obtained solutions of the transition temperature $t_c$ shown in Figs.~\ref{fig:SolutionsRTLG}(d)-\ref{fig:SolutionsRTLG}(f) are shown here in (d)-(f), respectively. Convergence of the numerically obtained solutions of the transition temperature $t_c$ shown in Figs.~\ref{fig:SolutionsRTLG}(g)-\ref{fig:SolutionsRTLG}(i) are shown here in (g)-(i), respectively. The numbers in different shapes (color coded according to $\mu$ values) correspond to the number of grid points ($N$) we divide the $(k_x, k_y)$ momentum space within the range $-0.25 \leq k_j \leq 0.25$ for $j=x$ and $y$ into after converting the integral over $\vec{k}$ in the gap equation into a discrete summation.  
}~\label{fig:ConvergenceRTLG}
\end{figure*}

At this point an important comment is due. Notice that in graphene heterostructure stacked in the rhombohedral fashion and subject to perpendicular electric displacement field, the effective Hamiltonian in terms of two low-energy sites can always be cast in the above form that is block diagonal in the valley and spin indices, while involving three Pauli matrices ($\beta_{1,2,3}$) in the sublattice or layer subspace. Concomitantly, following the discussion accounting for competing orders and systematic lifting of the spin and valley degeneracy, leading to the formation of half-metal and QM, is equally applicable in any rhombohedrally stacked graphene heterostructure (with an arbitrary number of layers), with our focus being on BBLG and RTLG here.

Now we consider the effect of the dominant channels of electronic interactions. Repulsive on-site Hubbard $U$ in graphene heterostructures induces a layer (or sublattice) antiferromagnet ordering with amplitude $\Delta_{\rm LAF}$, described by the matrix operator $\Delta_{\rm LAF} \Gamma_{303}$, where we arbitrarily (without any loss of generality) choose the spin quantization along the $z$ axis. Both the $V$ and $\Delta_{\rm LAF}$ terms introduce a spectral gap. However, due to the commutation of the corresponding matrix operators $[\Gamma_{003},\Gamma_{303}]=0$, the simultaneous presence of these orderings lifts the spin degeneracy, but maintains the valley degeneracy as both orders are valley un-polarized (appearing with the $\tau_0$ matrix). The next relevant component of short-range interactions is intralayer next-nearest-neighbor Hubbard repulsion, which in the spin-polarized system induces quantum anomalous Hall ordering with amplitude $\Delta_{\rm QAH}$, described by the operator $\Delta_{\rm QAH} \Gamma_{033}$~\cite{szaboroy:selectionrule}. As $[\Gamma_{033},\Gamma_{003}]=[\Gamma_{033},\Gamma_{303}]=0$, the Haldane mass represents yet another commuting order. Therefore, a simultaneous presence of these three orderings lifts the complete fourfold degeneracy of the normal-state band structure~\cite{RTLGrecent:6, BBLGrecent:2}. These outcomes can be appreciated straightforwardly by diagonalizing the effective single-particle Hamiltonian in the presence of layer-antiferromagnet only (for half-metal) and its simultaneous presence with the anomalous Hall ordering, given by 
\begin{eqnarray}~\label{eq:SPQMgeneral}
H^{j}_{\rm SP} (\Delta_{\rm LAF}, \Delta_{\rm QAH}) = H_{j} + \Delta_{\rm LAF} \Gamma_{303} + \Delta_{\rm QAH} \Gamma_{033} \nonumber \\
\end{eqnarray} 
for $j=$ BBLG and RTLG. In the two-fold valley degenerate half-metal phase, the effective single-particle Hamiltonian is given by $H^{j}_{\rm SP} (\Delta_{\rm LAF}, 0)$, and this phase is realized for the chemical potential  $\mu$ satisfies $V+\Delta_{\rm LAF}< \mu < |V-\Delta_{\rm LAF}|$. By contrast, Eq.~\eqref{eq:SPQMgeneral} describes a valley and spin non-degenerate QM when the chemical potential $\mu$ satisfies $|V-\Delta_{\rm LAF}+\Delta_{\rm QAH}|<\mu<|V-\Delta_{\rm LAF}-\Delta_{\rm QAH}|$, assuming $\Delta_{\rm LAF}>\Delta_{\rm QAH}$, which is natural as the strength of the on-site Hubbard repulsion is expected to be stronger than that for the intra-layer next-nearest-neighbor repulsion~\cite{katsnelson:hubbard}. Thus, in the QM phase the uniform mass is given by $m(\vec{k}=0)=m_0 \equiv |V-\Delta_{\rm LAF}-\Delta_{\rm QAH}|$. It can be immediately appreciated that the above discussion on the cascade of degeneracy lifting in rhombohedral graphene multi-layer leading to the formation of half-metal and QM is applicable to all such systems, where only the form of the non-interacting Hamiltonian changes as we show here explicitly by considering BBLG and RTLG in tandem. On the same token, one can immediately extend the discussion on PDW in the QM in all these systems, which thus far has only been observed in tetralayer, pentalayer, and hexalayer rhombohedral graphene~\cite{PDW:Exp1, PDW:Exp2, PDW:Exp3}. As mentioned in the Introduction, the sites constituting the low-energy description in rhombohedral-stacked graphene heterostructure with odd and even number of layers form an effective honeycomb and prismlike lattice, respectively, where the PDWs therefore take Kekul\'e and columnar shapes, shown in Figs.~\ref{fig:lattice}(d) and~\ref{fig:lattice}(e).

\section{Band parameters}~\label{append:bandparameters}

In this appendix, we mention various parameter values that have been used in our calculations. From experimental data~\cite{BBLG:bandparams, BBLG:Exp1, BBLG:Exp2, BBLG:Exp3, RTLG:Exp1} and \emph{ab initio} calculations~\cite{RTLG:bandparams}, we obtain the values of various band parameters and displacement electric field ($D$), pertinent to the QM phase in BBLG and RTLG. We set the overall energy scale in these two systems via the intralayer nearest-neighbor hopping amplitude $t\approx3.0\ {\rm eV}$, and measure the rest of the parameters in units of $t$. In BBLG, $t$, $t_\perp$, and $t_3$ are available from infrared spectroscopy measurements, yielding $t=3.0$ eV, $t_\perp=0.4$ eV, and $t_3=0.1$ eV~\cite{BBLG:bandparams}. On the other hand, the range of the displacement field $D$ for which the QM appears in BBLG can be read off in units of ${\rm V/nm}$ from the phase diagrams shown in Refs.~\cite{BBLG:Exp1, BBLG:Exp2, BBLG:Exp3}. The resulting potential energy $V$ (entering the Hamiltonian) can then be calculated as $V=D d$, where $d\approx 0.34\ {\rm nm}$ is the interlayer distance. Specifically, in BBLG the QM is observed for $D=0.3-0.7$ V/nm, yielding $V=0.102-0.238$ eV. All these parameter values, when expressed in terms of the corresponding dimensionless quantities, we find them to be $t_\perp/t=0.133$, $t_3/t=0.1$, and $V/t=0.034-0.080$. The chosen values of the (dimensionless) chemical potential $\mu$ ($0.0320 \leq \mu \leq 0.050$), used for the calculations (Figs.~\ref{fig:SCbandBBLG} and~\ref{fig:SolutionsBBLG}), are reasonably consistent with the range of $V$ over which QM has been observed in experiments, which in addition also includes the doping regimes that are slightly away from it.

The tight-binding parameters in RTLG are known from \emph{ab initio} calculations~\cite{RTLG:bandparams}, whereas the displacement field where the QM is observed can be read off from the experimentally observed phase diagram~\cite{RTLG:Exp1}. Note that in RTLG the middle layer is inert and the relevant bands are contributed by the top and bottom layers. Correspondingly, the potential energy in RTLG due to the displacement field is calculated as $V=D (2d)$, where $2d\approx 0.68\ {\rm nm}$. In RTLG, $t=3$ eV, $t_\perp=0.4$ eV, and $t_3=0.0171$ eV, yielding the dimensionless parameters $t_\perp/t =0.16$ and $t_3/t=0.006$. The QM phase in this system has been seen for $D=0.3-0.6$ V/nm and the corresponding $V=0.2-0.41$ eV, which in our calculation in parameterized by the range of a dimensionless quantity $V/t=0.067-0.0137$. The chosen values of the (dimensionless) chemical potential $\mu$ (Figs.~\ref{fig:SCbandRTLG} and~\ref{fig:SolutionsRTLG}) fall within this range of $V/t$. Finally, we note that although we have used various parameter values that are reasonably consistent with experiments and \emph{ab initio} calculations, our results are qualitatively insensitive to them, and depends only on the existence of a QM and trigonal warping in these systems.

\section{Band projection of PDW and emergent odd-parity pairing}~\label{append:banddiaginalization}

In this appendix, we display the technical details associated with the band diagonalization procedure within the QM in BBLG and RTLG, and the extraction of the intraband component of the PDW in these systems. The starting point is the single-particle Hamiltonian that describes the QM in these systems, which in the Nambu-doubled basis takes the generic block diagonal form 
\begin{eqnarray}
H^{\rm Nam}_{\rm QM} = \left(\begin{array}{cc}
{\boldsymbol \beta} \cdot \vec{d}_p (\vec{k}) & {\boldsymbol 0} \\
{\boldsymbol 0} &  -{\boldsymbol \beta} \cdot \vec{d}_h (\vec{k})
\end{array}
\right),
\end{eqnarray}
where ${\boldsymbol 0}$ is a two-dimensional null matrix, ${\boldsymbol \beta}=(\beta_1, \beta_2, \beta_3)$ and $\vec{d}_j (\vec{k})=(\vec{d}^1_j (\vec{k}), \vec{d}^2_j (\vec{k}), \vec{d}^3_j (\vec{k}))$ for $j=p,h$. The explicit forms of $\vec{d}_j (\vec{k})$ in BBLG and RTLG can be obtained from Eqs.~\eqref{eq:QMBBLGHamil},~\eqref{eq:QMRTLGHamil}, and~\eqref{eq:HNambu}, which we do not show here explicitly for brevity. Due to the block-diagonal form of $H^{\rm Nam}_{\rm QM}$ the unitary operator ($U_{\rm diag}$) that diagonalizes it also assumes the block-diagonal form $U_{\rm diag}=U^{p}_{\rm diag} \oplus U^{h}_{\rm diag}$, where  
\begin{eqnarray}
U^{j}_{\rm diag}= \frac{1}{\sqrt{2 E_j}}\left( \begin{array}{cc}
\frac{d^1_j-i d^2_j}{[E_j - d^3_j]^{1/2}} & - \frac{d^1_j-i d^2_j}{[E_j + d^3_j]^{1/2}} \\
{[E_j-d^3_j]^{1/2}} & [E_j + d^3_j]^{1/2}
\end{array}
\right)
\end{eqnarray}
for $j=p,h$. Here, we have used $d^\ell_j \equiv d^\ell_j(\vec{k})$ for brevity with $\ell=1,2,3$. Expression for $E_{p,h}$ for BBLG and RTLG are shown in Eqs.~\eqref{eq:EphBBLG} and~\eqref{eq:EphRTLG}, respectively. After the unitary rotation by $U_{\rm diag}$, the Hamiltonian $H^{\rm Nam}_{\rm QM}$ becomes diagonal, taking the explicit form
\begin{eqnarray}
U^\dagger_{\rm diag} H^{\rm Nam}_{\rm QM} U_{\rm diag}= {\rm Diag}. \left(E_p, -E_p, -E_h, E_h \right). 
\end{eqnarray}   
Isolating the contributions for the conduction band (CB) and taking into account finite chemical potential, we arrive at the single-particle Hamiltonian for the QM in the band basis as follows
\begin{eqnarray}~\label{eq:CBHamil}
H^{\rm Nam, CB}_{\rm QM} &=& {\rm Diag}. \left( E_p-\mu, - (E_h-\mu)\right) \nonumber \\
&=& \eta_0 E_- \eta_0 + \left( E_+ -\mu \right) \eta_3,
\end{eqnarray}
where $E_\pm=(E_p \pm E_h)/2$ and the set of Pauli matrices $\{ \eta_\nu\}$ operates on the Nambu or particle-hole indices.

Next, we act with the unitary operator $U_{\rm diag}$ on the pairing term ($H_{\rm pair}$) shown in Eq.~\eqref{eq:HNambu}. Schematically, this procedure leads to 
\begin{eqnarray}
U^\dagger_{\rm diag} H_{\rm pair} U_{\rm diag} &=& U^\dagger_{\rm diag} \Delta(\eta_1 \cos\phi+\eta_2 \sin\phi) M U_{\rm diag} \nonumber \\
&=& \Delta (\eta_1 \cos\phi+\eta_2 \sin\phi) 
\left( \begin{array}{cc}
a & b \\
c & d
\end{array}
\right),
\end{eqnarray}
where the quantity $a$ ($c$) represents the intraband component of the pairing within the conduction (valence) band, $b$ and $d$ capture its interband components, and in the above expression $M=\beta_2$. From now onward we solely focus on $a$, given by 
\begin{eqnarray}~\label{eq:intrabandpair}
a &=& \frac{i}{2\sqrt{E_p E_h}} \bigg[ \sqrt{\frac{E_p-d^3_p}{E_h-d^3_h}} (d^1_h -i d^2_h) \nonumber \\
&-&\sqrt{\frac{E_h-d^3_h}{E_p-d^3_p}} (d^1_p -i d^2_p)  \bigg],
\end{eqnarray}
where we set $\vec{d}_j(\vec{k}) \equiv \vec{d}_j$ for brevity and without any loss of generality set the superconducting phase $\phi=0$.

We first consider BBLG, for which the readers can immediately verify that if we set the trigonal warping $t_3=0$, then $a=0$. This is consistent with the facts that pairing among non-degenerate fermionic degrees of freedom can only take place in the odd-parity channel to ensure the overall antisymmetric property of the Cooper pair wave functions and in BBLG only trigonal warping (also described by odd-parity functions) can source such odd-parity component to any local pairing term (Kekul\'e PDW in this case). Identifying trigonal warping as the source of odd-parity pairing, we now retain only the term that is rotationally invariant as it yields maximal gain in condensation energy and highest transition temperature, and nucleation of this component reduces the propensity of condensation of other components by maximally gapping out the Fermi ring. Finally, since in the presence of a Fermi ring the pairing interaction dominantly exists near it, we set $E_{p,h} \approx \mu$ in the prefactor residing outside the square brackets in Eq.~\eqref{eq:intrabandpair}. With this realistic simplification, we obtain 
\begin{equation}
a \approx \Delta \; \frac{\alpha_1}{\mu} \; \left( \eta_1 p_1(\vec{k}) - \eta_2 p_2(\vec{k})\right),
\end{equation}  
which when combined with $H^{\rm Nam, CB}_{\rm QM}$ [see Eq.~\eqref{eq:CBHamil}] yields the total reduced BCS Hamiltonian shown in Eq.~\eqref{eq:ReducedBCSBBLG}, which assumes the form of a topological odd-parity $p+ip$ pairing, as the odd-parity nature of the pairing is ensured by the trigonal warping which is also described in terms of two odd-parity $p$-wave harmonics.

In RTLG, the intraband component of the columnar PDW takes the schematic form shown in Eq.~\eqref{eq:intrabandpair}, which upon setting the amplitude of the trigonal warping $t_3=0$, yields (along with the simplification $E_{p,h} \approx \mu$)
\begin{equation}
a \approx \Delta \frac{\alpha_3}{\mu} \; \left( \eta_1 f_1(\vec{k}) - \eta_2 f_2(\vec{k})\right).
\end{equation}    
In this case, the requisite odd-parity component to the local pairing term is sourced by the dominant cubic dispersion in the normal state, and the corresponding total reduced BCS Hamiltonian [see Eq.~\eqref{eq:ReducedBCSRTLG}] assumes the form of a topological odd-parity $f+if$ pairing. We note that both the emergent $p+ip$ (in BBLG) and $f+if$ (in RTLG) pairings, resulting from the corresponding PDW belong to class D in ten-fold Altland-Zirnbauer classification scheme~\cite{AZ:1, AZ:2}.

\section{Convergence of numerical solutions}~\label{append:convergence}

In this Appendix, we comment on the convergence of the numerical solutions of the dimensionless pairing amplitude ($\Delta$) at zero temperature and transition temperature ($t_c$), shown in Fig.~\ref{fig:SolutionsBBLG} (for BBLG) and Fig.~\ref{fig:SolutionsRTLG} (for RTLG), obtained from the self-consistent gap equation, discussed in Sec.~\ref{Sec:meanfield}. These analyses are shown in Fig.~\ref{fig:ConvergenceBBLG} for BBLG and Fig.~\ref{fig:ConvergenceRTLG} for RTLG. As mentioned in Sec.~\ref{Sec:meanfield}, instead of performing numerical integrations over the momentum $\vec{k}$, we covert them into discrete summations and divide the momentum range in the $(k_x,k_y)$ plane into $N$ number of equi-spaced grids. For each value of the chemical potential ($\mu$), the numerical summations are performed for three different values of $N$ that are largely apart from which we test the convergence of the numerical solutions. The values of $\mu$ and $N$ are quoted in the legends of Figs.~\ref{fig:ConvergenceBBLG} and~\ref{fig:ConvergenceRTLG}. For each topology of the underlying Fermi rings (disjoint, annular, and simply connected), we choose three representative values for $\mu$ in BBLG and RTLG. Typically, we choose $N=50000$, $100000$, and $150000$. The self-consistent solutions for $t_c$ for all chosen parameter values show excellent convergence in BBLG and RTLG. Seldom, the solutions for $\Delta$ at these specific values of $N$ show minor numerical instabilities for extremely small values of (dimensionless) $\Delta$ in the range $10^{-6}<\Delta < 10^{-5}$. We then choose slightly different values of $N$. We also notice that such numerical instabilities disappear immediately as soon as we introduce a small temperature $\sim 10^{-7}$ due to the thermal damping factors. Still, we choose to present the solutions of $\Delta$ exactly at zero temperature.


\end{document}